%% file: cognition.tex
\documentclass[12pt,openany]{report}

\usepackage{biblatex}
\usepackage{amsmath}
\usepackage{amssymb}
\usepackage{mathtools}
\usepackage{mathrsfs}
\newcommand{\F}{\mathcal{F}}
\newcommand{\R}{\mathbb{R}}
\newcommand{\N}{\mathbb{N}}

\DeclareMathOperator*{\argmax}{arg\,max}

\usepackage{amsthm}
\newtheorem{theorem}{Theorem}[chapter]
\theoremstyle{definition}
\newtheorem{define}{Definition}[chapter]

\usepackage{tikz}
\tikzset{>=latex}
\tikzset{graph/.pic={
  \node (a) at (0,3)      [circle,draw] {a};
  \node (b) at (2,3.5)    [circle,draw] {b};
  \node (c) at (1,1.5)    [circle,draw] {c};
  \node (d) at (3.5,2)    [circle,draw] {d};
  \node (e) at (0,0)      [circle,draw] {e};
  \node (f) at (2.25,0.5) [circle,draw] {f};
  \node (g) at (4,0)      [circle,draw] {g};
  \node (h) at (1,-1.5)   [circle,draw] {h};
  \node (i) at (3,-1.5)   [circle,draw] {i};
  \draw [->] (a) to (b);
  \draw [->] (a) to [bend right=30] (c);
  \draw [->] (b) to [bend right=30] (d);
  \draw [->] (b) to (c);
  \draw [->] (c) to [bend right=30] (a);
  \draw [->] (c) to (d);
  \draw [->] (c) to (f);
  \draw [->] (d) to [bend right=30] (b);
  \draw [->] (d) to (g);
  \draw [->] (e) to (f);
  \draw [->] (e) to (c);
  \draw [->] (e) to (h);
  \draw [->] (f) to [bend right=30] (h);
  \draw [->] (f) to (i);
  \draw [->] (f) to (d);
  \draw [->] (f) to (g);
  \draw [->] (h) to [bend right=30] (f);
  \draw [->] (i) to (g);
}}

\usepackage{tocloft}
\usepackage[toc,page]{appendix}
\usepackage{subfiles}
\usepackage{comment}
\usepackage{float}
\usepackage[shortlabels,inline]{enumitem}
\usepackage{graphicx}
\usepackage{outlines}

\usepackage{indentfirst}
\usepackage{setspace}
\title{Cognition in Dynamical Systems \\ {\large Second Edition}}
\author{Jack Hall}

\begin{document}

\maketitle

\begin{abstract}
  \subfile{abstract.tex}
\end{abstract}

\pagenumbering{gobble}
\subfile{acknowledgements.tex}

\setlength{\cftbeforetoctitleskip}{20pt}
\renewcommand{\cftchappresnum}{Chapter }
\setlength{\cftchapnumwidth}{6em}
\tableofcontents

\chapter{Introduction}
\label{chp:intro}

\pagenumbering{arabic}

\subfile{lead.tex}
\subfile{overview.tex}

\chapter{Signal Propagation}
\label{chp:propagation}

\subfile{signaling_intro.tex}
\subfile{naive.tex}
\subfile{dynamics.tex}
\subfile{stability.tex}
\subfile{continuity.tex}
\subfile{signaling_summary.tex}

\chapter{Pattern and Feedback}
\label{chp:pattern}

\subfile{pattern_intro.tex}
\subfile{existing.tex}
\subfile{emergence.tex}
\subfile{specifications.tex}

\chapter{Applications}
\label{chp:applications}

\subfile{applications.tex}
\subfile{autocatalysis.tex}
\subfile{neural_nets.tex}
\subfile{markets.tex}
\subfile{ant_colonies.tex}

\chapter{Conclusions}

\subfile{meta.tex}
\subfile{review.tex}
\subfile{future.tex}


\begin{singlespace}
  \printbibliography[heading=bibintoc,title=References]
\end{singlespace}

\end{document}

%% file: abstract.tex

Cognition is the process of knowing.
As carried out by a dynamical system, it is the process by which the system absorbs information into its state.
A complex network of agents cognizes knowledge about its environment, internal dynamics and initial state by forming emergent, macro-level patterns.
Such patterns require each agent to find its place while partially aware of the whole pattern.
Such partial awareness can be achieved by separating the system dynamics into two parts by timescale: the propagation dynamics and the pattern dynamics.
The fast propagation dynamics describe the spread of signals across the network.
If they converge to a fixed point for any quasi-static state of the slow pattern dynamics, that fixed point represents an aggregate of macro-level information.
On longer timescales, agents coordinate via positive feedback to form patterns, which are defined using closed walks in the graph of agents.
Patterns can be coherent, in that every part of the pattern depends on every other part for context.
Coherent patterns are acausal, in that
\begin{enumerate*}[label={\alph*)}]
  \item they cannot be predicted and
  \item no part of the stored knowledge can be mapped to any part of the pattern, or vice versa.
\end{enumerate*}
A cognitive network's knowledge is encoded or embodied by the selection of patterns which emerge.
The theory of cognition summarized here can model autocatalytic reaction-diffusion systems, artificial neural networks, market economies and ant colony optimization, among many other real and virtual systems.
This theory suggests a new understanding of complexity as a lattice of contexts rather than a single measure.

%% file: acknowledgements.tex

\begin{center}
  \topskip0pt
  \vspace*{\fill}

  \textbf{Acknowledgments}

  \bigskip
  Many professors have been very patient with me. \\
  Dawn Tilbury and Benito Fernandez set me on my path. \\
  Raul Longoria kept me in graduate school. \\
  Matthew Campbell and Maxwell Stinchcombe made me explain myself. \\
  Luis Sentis balanced my realism with optimism.

  \bigskip
  My friends were a boon. \\
  Courtney Shell and Nick Paine kept me sane. \\
  Felipe Lopez, Prashant Rao and Taylor Niehues kept me laughing and made sure my glass was full. \\
  Gray Thomas always listened to my latest strange idea.

  \bigskip
  My mother and sisters held me up and reminded me that they love me anyway.

  \bigskip
  Thank you all.

  \vspace*{\fill}
\end{center}
\pagebreak

%% file: lead.tex

Ants are both familiar and mysterious.
You can find a colony almost anywhere, but how do they dig such intricate tunnels, forage as a group and wage war?
How do they decide which of these they should all be doing, when each ant - including the queen - is quite stupid?
Ants signal each other by touch, sound and pheremone, and these signals coordinate them in ways no individual ant could fathom.
The signals propagate from ant to ant, inhibiting, modifying, amplifying and directing each other so that the ants, in obeying them, act for the colony's good.

This complexity would not be necessary in a simple or isolated environment.
A colony must simultaneously find enough food, secure its nest against predators, dispose of waste and care for its young.
Despite their caste system most ants perform many tasks.
A worker may retrieve some food, store it, repair a collapsed tunnel, then go back to harvesting food.
Each action is a response to stimulus from the environment and from other ants.
An ant that signaled for more food may become sidetracked to help the queen brood, but the hunger signal remains and propagates to other ants, some of whom respond.
The colony as a whole constantly responds to myriad stimuli, allocating resources and collectively making decisions.

Asked to explain this collective cognition, you could point to evolution by natural selection.
Every generation, each ant's response to stimulus changed slightly, which tweaked the collective behavior.
Successful tweaks survived, and ant societies grew more complex.
But this story tells us little about how each ant colony reacts to its environment.
How do those entangled patterns of signaling bridge environmental stimulus to colony survival?
How might they be engineered?

\begin{quote}
  \textbf{Cognition} is the process of knowing.
\end{quote}
While cognition commonly refers only to human thought, this definition comes from cybernetics.
For a dynamical system to \emph{know}, information must be encoded in its state.
That information may come from the environment or quirks of the system's own nature.
\begin{quote}
  A system \textbf{cognizes} when it absorbes information, internal or external, into its state.
\end{quote}

To store any significant amount of information, the system must have a very ``large'' state space.
Real numbers can technically hold an arbitrary amount of information, and the theory of dynamical systems whose state are vectors of reals is well-developed.
Could we use that theory to describe cognition?
Unfortunately, no.
To store a message, the system must find a corresponding region of its state space and remain there. 
The system's dynamics function as an encoder of the message.
To increase the size of the message, a state space would need to be split into increasingly small stable regions, and the dynamics would become increasingly difficult to analyze.
Furthermore, each such region would border on a limited number of others, making it hard for the system to encode an arbitrary message starting from arbitrary initial conditions.\footnote{
If the system starts with no prior information about the message, the initial conditions will certainly be arbitrary.}

Alternatively, we can increase the dimensionality of the state space.
To keep the dynamics tractable, we need to restrict the ways in which the state variables affect each other.
Complex networks fit the bill.
The state space of a complex network has an arbitrary number of dimensions, and each agent affects only itself and its neighbors.
Certain complex networks are already known to absorb information from their environment.\footnote{
This is usually described as acquiring order from the environment.}
All living systems are complex networks, as are some nonliving systems.
\begin{quote}
  A \textbf{complex network} is a system of agents that selectively interact with each other in ways that are hard to understand solely from knowledge of the individual agents.
\end{quote}
\begin{quote}
  An \textbf{agent} is an indivisible entity which acts of its own accord.
\end{quote}

If it was not already clear, I will not investigate ant colonies in particular.
Instead, I will look at an entire class of complex networks - to which ant colonies belong - that produce unpredictable but stable and coherent patterns.
Each of these words - unpredictable, stable, coherent - is both vital and frustratingly vague.
If the patterns were predictable, we would have no use for the complex system.\label{unpredictable}
If the system did not find a stable region of its state space, then it cannot store information.\label{stable}
And to store a single large message rather than many unrelated and trivial messages, agents must specialize and coordinate - even though each agent cannot directly affect most others.
\begin{quote}
  A pattern is \textbf{coherent} if each part of it depends on the rest.
  \label{coherent}
\end{quote}

Unfortunately, no mathematics exist that describe pattern or stability in a complex system.\cite{manson}
Without such mathematics, any reasoning about complex systems will be as fuzzy as the words I defined in the last paragraph.
I will articulate, in this thesis, mathematics that underpin cognition.
These mathematics could be used to:
\begin{itemize}[noitemsep]
  \item control large distributed systems like smart grids and reconfigurable factories
  \item solve hard optimization problems and automate engineering design
  \item coordinate and plan complicated robotic motions
\end{itemize}
The mathematics of cognition will be necessary - though likely not sufficient - for growing strong artificial intelligence.
We may also come to better understand natural systems like market economies and ecosystems.
That understanding could help us to develop better public policy and safely manage the environment.
And there are many other ways to apply cognition; I am sure that I will never see them all.

%% file: overview.tex

The next two chapters will construct this new theory.
I will begin with the interactions between agents, studying the way information propagates across a network.
Then, in Chapter \ref{chp:pattern}, I will set forth how agents coordinate and form patterns through the use of positive feedback.
In Chapter \ref{chp:applications}, I will apply the mathematics of the first two chapters to a set of more concrete models.
My goal is a set of abstractions that explain a wide range of concrete cognitive systems.

In my last chapter, I will examine why I chose those abstractions and not others.
The concept of context will prove to be a strong guiding principle.
It may seem, by the end, that I present some of this material out of order.
I do apologize for any difficulties in understanding my text, but we are all at the mercy of cognition itself.
It emerges whole, of its own accord.

%% file: signaling_intro.tex

This chapter will describe how information propagates across a complex network.
By the end, I will have explained how agents gain access to macro-level information through the aggregations each agent performs for its neighbors.

%% file: naive.tex

\section{A Naive Approach}
\label{sec:naive}

The standard model of a complex network is a directed graph $G=(V,E)$ where the vertices $V$ are the agents.
For convenience, I will define two functions, $T(v) = \{(v',v) : (v',v) \in E\}$ and $F(v) = \{(v,v') : (v,v') \in E\}$, to give the arcs incident to and incident from agent $v$, respectively.
If there is no chance of ambiguity, these can be written $Tv$ and $Fv$.
The state of each agent $v$ is governed by some flow function $\psi_v^t$.
The graph specifies which agents directly affect the state of which others; the state of agent $v$ depends on the state of agent $v'$ if and only if $(v',v) \in E$.
This means that if time is discrete ($t \in \mathbb{Z}$), then $\psi_v^1$ maps the states of agents $\{v\} \cap Tv$ to the state of agent $v$, and if time is continuous ($t \in \R$), then $d \psi_v / dt$ is a function of those same agent states.
By changing their states over time and influencing their neighbors, agents are said to \emph{act} per their definition.

\begin{figure}
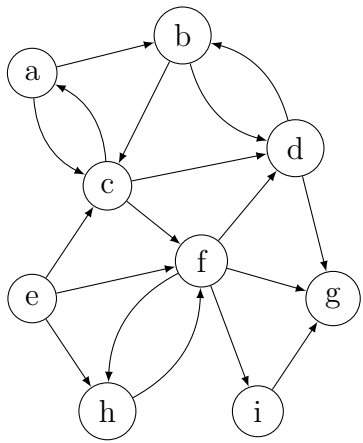

  \centering
  \tikz \pic at (0,0) {graph};
  \caption[A directed graph]{A directed graph. As an example of my notation, $Tb = \{(a,b),(d,b)\}$ and $Fb = \{(b,c), (b,d)\}$.}
\end{figure}

For agents to coordinate with each other, they must exchange information.
Unfortunately, we cannot study solely how information spreads across the network, because the standard model ties up that spread with changes in agent state.
To separate the two, I will complicate the model slightly.
\begin{quote}
  A signal is the action of one agent on another.
\end{quote}
Let every signal be a point in $X$.
The signal $x_e \in X$ moves across arc $e \in E$.
The vector of signals in a set of arcs $A \subseteq E$ is a point $x_A \in X^{|A|}$, the $|\cdot|$ notation being the cardinality of $A$.
I will use different subscripts on the same variable to select subvectors, so $x_A$ is a subvector of $x_E \in X^{|E|}$, and $z_A$ is a subvector of $z_E \in X^{|E|}$.
Agent $v$'s state now depends the signals $x_{Tv}$ rather than the states of agents in ${v': (v',v) \in Tv}$.
I will take the unusual step of identifying each agent's state with a function that maps input signals to output signals.
\begin{equation}
  \label{eqn:agent_function}
  f_v: X^{|Tv|} \mapsto X^{|Fv|} \qquad f_v \in \F_v
\end{equation}
$\F_v$ is the state space through which each agent $v$ moves. 
Having now adjusted the standard model, how do agents exchange information?

I will briefly take the role of a single agent.
If I am to join a coherent pattern, I must coordinate with other agents to which I am not necessarily adjacent.
But I am only a simple agent, so my state cannot specifically reflect these distant agents.\footnote{
  I use ``simple'' here in a very narrow sense.
  An agent can itself be complex, but by this assumption it cannot cognize the whole network of which it is a part.}
Any information I exchange with distant agents must therefore flow through my neighbors.
I can tell my neighbors apart, but there is no way to know the ultimate source of their information - or for them to know the ultimate destination of mine.
Nonetheless the information channel between me and any distant agent consists of chains of adjacent agents.
All agents constantly move, which changes the way they direct my signals.
By the time my signal reaches a distant agent, and that agent responds, the information channel between us has changed in response to other signals as well as its own collective state.
My state has changed too, for that matter.

You might argue that agent movements in and of themselves do not doom my communication with the distant agent.
If agents move cyclically, then I may be able to sychronize to some extent with that distant agent.
But if agents only move in cycles, the whole network will simply vibrate indefinitely, making no progress towards a coherent pattern.
Insofar as a cycle is steady, it amounts to a fixed state.
And if the cycle changes over time (again at a rate similar to that of signal travel), then our information channel again loses integrity.

So coordinating with distant agents seems unlikely.
But how then can they form the coherent patterns necessary to store large messages?

%% file: dynamics.tex

\section{Propagation Dynamics}
\label{sec:aggregation}

To foster coherent patterns, let us try slowing down agent movement to the point where they are quasi-static on signaling timescales.
Each agent, if its function is fixed, simply maps input signals to output signals.
Our single complex system is now two interrelated systems: one for the propagation of signals across the network and one for changes in agent state.
For brevity, let $\F_V = \prod_{v \in V} \F_v$.
A point $f_V \in \F_V$ is a vector of functions $f_v$ for every agent $v \in V$.
Similar to the convention of signal notation, $f_v$ is a component of $f_V$, and $g_v$ is a component of $g_V$.

\begin{define}
  \label{dfn:propagation}
  The \textbf{propagation dynamics} on $G=(V,E)$ are a flow (or an iterated function) $\Gamma^t: \F_V \times X^{|E|} \mapsto X^{|E|}$ parameterized by the agent functions $f_V$.
  A point $z_E$ is a fixed point of the propagation dynamics only if $z_{Fv} = f_v(z_{Tv})$ for all agents $v \in V$.
\end{define}

The propagation dynamics define the way information spreads across the network.
Usually they are stochastic and discrete, although as you will see in Chapter \ref{chp:applications}, the propagation dynamics can be stochastic or deterministic, discrete or continuous.

Our second system - the slow counterpart to the propagation dynamics - describes the emergence of patterns.
To limit the speed of agent motion, there must be a metric on $\F_v$.
I will call it $d_v$. \label{agent_metric}

\begin{define}
  \label{dfn:pattern_dynamics}
  The \textbf{pattern dynamics} on $G = (V,E)$ are a continuous flow $\psi_V^t: \F_V \mapsto \F_V$ that describes the motion of every agent function over time, starting from any point in $\F_V$.
  The motion of a single agent $v$ is denoted $\psi_v^t$ and is Lipschitz continuous (with a Lipschitz constant $L \geq 0$):
  \begin{equation}
    d_v \big( f_v, \psi_v^t(f_v) \big) \leq L |t|
    \qquad t \in \R, \quad \forall f_v \in \F_v, \quad \forall v \in V
  \end{equation}
  An agent's motion at any instant is determined only by its current state and the signals it receives.
\end{define}

I would like to make the last statement in that definition more formal, but I cannot do so quite yet.
If $\psi_v$ were differentiable, then its time derivative would be a function of the current $f_v$ and current $x_{Tv}$.
Unfortunately, requiring differentiability would cripple a theory of cognition; there are many cognitive networks with non-differentiable states.
I will discuss several of them in Chapter \ref{chp:applications}.
For now, think of each continuous agent flow $\psi_v$ as approximated by an iterated function of the current state $f_v$ and the current input signals $x_{Tv}$, where each iteration moves the agent infintesimally.

The propagation and pattern dynamics must be kept separate.
But what does this mean?
Say that the propagation dynamics always converge to a fixed point.
According to Definition \ref{dfn:propagation}, the fixed point depends on the agent functions.
Let $p:\F_V \mapsto X^{|E|}$ map the agent functions to a fixed point of the corresponding propagation dynamics.
As the pattern dynamics progress, the fixed point of the propagation dynamics would move.
To be separate from the pattern dynamics, the propagation dynamics should stay arbitrarily close to their fixed point.
That way, each agent $v$ will change its state as a function of $p(f_V)_{Tv}$.

``Closeness'' implies metric on $X^{|E|}$.
Let $(X, \rho)$ be a complete metric space and $A \subseteq E$.
Define a metric $\rho_A(x_A, y_A) = \sum_{e \in A} \rho(x_e, y_e)$ on $X^{|A|}$.
$(X^{|E|}, \rho_E)$ is also a complete metric space.\footnote{
  I have chosen the L1 norm to combine vectors of signals, but there are other options.}
Resist the temptation to think of signals as real numbers.
This interpretation is quite limiting, as you will see in Chapter \ref{chp:applications}.
A better interpretation of signals is as measureable sets of stimuli.
If $\mu: X \mapsto [0, \infty)$ is the measure and $\triangle$ is the symmetric difference, then $\rho(x,y) = \mu(x \triangle y)$.
\label{set_signals}

\begin{define}
  \label{dfn:separation}
  At time $t=0$, let $x_E$ be the state of the propagation dynamics and $f_V$ be the state of the pattern dynamics.
  The pattern dynamics are \textbf{separated} from the propagation dynamics at time $t$ if there exists an arbitrarily small $\epsilon>0$ such that:
  \begin{equation}
    \rho_E \Big( \Gamma^t \big(\psi_V^t(f_V), x_E \big), p \big( \psi_V^t(f_V) \big) \Big) \leq \epsilon
  \end{equation}
\end{define}

Agent states map to signals via $p$, which in turn determine how those agent states change over time.
I will provide an example proof of separation in Section \ref{sec:continuity}.

The propagation dynamics, separated from the pattern dynamics, solve a major standing problem in complex systems.
Emergence is generally agreed to come from feedback loops between the micro and macro levels of a complex system\cite{dewolf_holvoet}, but most models have trouble conveying macro-level information to individual agents.
They mostly choose between two approaches.\footnote{
  The few exceptions - artifical neural networks come to mind - invariably mimic specific natural systems.
  No one has explained precisely why they work.}
The first - taken by cellular automata - is to let information percolate on its own.
As we saw in Section \ref{sec:naive}, this is no different from ignoring the problem and hoping for a miracle.

The second approach is to compute a single, crude aggregate of all the agent states and make that single aggregate available to every agent.
These crude aggregations force a tradeoff between agent complexity and pattern coherence.
On one end of the scale, the crude aggregate perfectly summarizes the state of the whole system.
Each supposedly simple agent must then be complex enough to parse this information and understand its part in the pattern - a clear contradiction.
Multi-agent systems usually operate at or near this extreme.
On the other end of the scale, the aggregate contains no useful information about macro-level pattern.
Without it, agents cannot specialize.
Without specialization, a pattern cannot be coherent.
Agent-based models often use these uninformative aggregations.

Propagation dynamics sidestep the complexity/coherence tradeoff by distributing the work of aggregation.
Each agent aggregates information for its neighbors.
This concept is not new; Freidrich Hayek had noticed by 1945 that actors in a market aggregate information about the scarcity of each product by setting prices.\cite{hayek}
To know how much of a good to make, the producers of that good need not trace all the ways in which it is used or transformed by the rest of the economy; they need only observe changes in price.
Likewise to know how much of each good to use, consumers need not keep track of improvements or disruptions in every supply chain.
No human could ever be fully aware of all the decisions made in the economy, yet those decisions cohere.

\begin{define}
  \label{dfn:aggregate}
  The \textbf{aggregate} of the graph $G$ and agent states $f_V$ is given by a function $p: \F_V \mapsto X^{|E|}$ such that $p(f_V)$ is always a fixed point of the propagation dynamics on $G$.
  The function $p$ is the \textbf{aggregation}.
\end{define}

In the course of modeling any particular cognitive network, it's natural to ask which information propagates and which doesn't.
To keep my theory as general as possible, I have avoided this question and tried not to assume anything about the answer.

%% file: stability.tex

\section{Existence of an Aggregation}
\label{sec:stability}

The aggregation $p$ only exists if the propagation dynamics always have a stable fixed point, preferably a unique one.
Any proof of convergence to a fixed point will depend on the particular application of this theory.
This section contains two simple examples.

Consider first the following propagation dynamics: every agent $v$, simultaneously and at intervals in time, applies $f_v$ to its inputs $x_{Tv}$ and updates its outputs $x_{Fv}$ with the result.
\begin{equation}
  \label{eqn:synchronous}
  \Gamma^1(f_V, x_E) = \big( f_v(x_{Tv}) \big)_{v \in V}
\end{equation}

\begin{define}
  Let $\alpha \in (0,1)$ and $x_{Tv}, y_{Tv} \in X^{|Tv|}$.
  An agent function $f_v$ \textbf{contracts} if:
  \begin{equation}
    \alpha \rho_{Tv}(x_{Tv}, y_{Tv}) \geq
    \rho_{Fv} \big( f_v(x_{Tv}), f_v(y_{Tv}) \big)
  \end{equation}
\end{define}

\begin{theorem}
  \label{thm:synchronous}
  If every agent function contracts, then the propagation dynamics defined by Equation \ref{eqn:synchronous} converge exponentially to unique fixed point.
\end{theorem}
\begin{proof}
  \begin{align*}
    \rho_E( \Gamma^1(f_V, x_E), \Gamma^1(f_V, y_E) )
    &= \sum_{e \in E} \rho \big( \Gamma^1(f_V, x_E)_e, \Gamma^1(f_V, y_E)_e \big) \\
    &= \sum_{v \in V} \sum_{j \in Fv} \rho \big( f_v(x_{Tv})_j, f_v(y_{Tv})_j \big) \\
    &= \sum_{v \in V} \rho_{Fv} \big( f_v(x_{Tv}), f_v(y_{Tv}) \big) \\
    &\leq \sum_{v \in V} \alpha \rho_{Tv}(x_{Tv}, y_{Tv})
  \end{align*}
  \begin{equation}
    \rho_E \big( \Gamma^1(f_V, x_E), \Gamma^1(f_V, y_E) \big)
    \leq \alpha \rho_E(x_E, y_E)
  \end{equation}
  $\Gamma^1$ is a contraction mapping.
  All contraction mappings converge exponentially to a unique fixed point.
\end{proof}

Updating all agents simultaneously requires the agents to synchronize, which - according to my assumption that agents only coordinate via signals - they are unable to do unless the propagation dynamics converge.
The synchronous propagation dynamics in Equation \ref{eqn:synchronous} are physically unlikely.

Consider now the propagation dynamics in which each agent updates randomly according to a Poisson process in time.
The Poisson processes are independent and identically distributed with a rate $\lambda$.
These asynchronous propagation dynamics $\Lambda$ are a jump process on $X^{|E|}$ whose jumps are themselves Poisson-distributed in time, albeit with a much faster rate than $\lambda$.
Letting $x_A \| x_B = x_{A \cup B}$ concatenate two vectors indexed by disjoint sets $A$ and $B$, each jump moves $x_E$ to $x_{E-Fv} \| f_v(x_{Tv})$ for a randomly drawn agent $v$.

\begin{theorem}
  \label{thm:asynchronous}
  If every agent function contracts, $\Lambda$ contracts to a unique fixed point.
\end{theorem}
\begin{proof}
  Since $\Gamma$ has only one fixed point $z_E$, then by Definition \ref{dfn:propagation} so does $\Lambda$ have the same fixed point.
  Let the interval $[a,b)$ be an interval in time over which every agent has updated at least once.
  The expected length of this interval can be derived from the Poisson distribution.
  \begin{align*}
    1/2 &= (1 - P(N(b-a)=0))^{|V|} \\
    1/2 &= (1 - e^{-\lambda (b-a)})^{|V|}
  \end{align*}
  \begin{equation}
    E(b-a) = -\frac{1}{\lambda} \ln(1 - 2^{-1/|V|})
  \end{equation}
  Because this interval grows only logarithmically with the number of agents, the number of agents matters little.
  The actual size of each interval has a probability distribution, but the law of large numbers asserts that the time-average of interval lengths gets arbitrarily close to the expectation of that length as time goes on.

  The propagation dynamics contract at least as much over the interval $[a,b)$ as they would in one synchronous update. 
  The distance between the trajectories from $x_E$ and $y_E$ over time is therefore bounded from above by an exponential decay.
  \begin{equation}
    \label{eqn:propagation_decay}
    \alpha^{t/(b-a)} \rho_E(x_E, y_E) \geq
    \rho_E \left(\Lambda^t(f_V, x_E), \Lambda^t(f_V, y_E) \right)
  \end{equation}
\end{proof}

Having shown that aggregations exist for propagation dynamics $\Gamma$ and $\Lambda$, I now ask: How complete is the aggregate?
For two distant agents to coordinate, the aggregate available to each must be sensitive to the state of the other.
In other words, the signals available to each agent should include at least \emph{some} information about every other agent.
Given the restriction on fixed points in Definition \ref{dfn:propagation}, this property depends only on the agent functions (and the graph) rather than propagation dynamics they parameterize.
I can see two scenarios in which information fails to spread properly.

If the graph grows too large and sparse, the shortest path between a pair of agents can grow arbitrarily long.
Agent contraction does not just bound the time in which signals converge, it also limits how far they can spread through the graph.
Agents can (and usually will) prioritize certain signals over others, which will cause the former to propagate further than the latter.
Nevertheless, contractive agents cannot pass on more than a fraction of the signals they receive.
Over a long enough path, even the strongest and most important of signals must fade. 
One solution is to give large networks the small-world property\footnote{
  in which the number of arcs to and from any particular agent has a geometric probability distribution.
  In simple terms: while most agents have relatively few arcs, a few agents have many, and a very few have a huge number.},
thereby putting an upper bound on the shortest path between any two agents.
Another solution is to relax the rate of convergence from exponential to asymptotic.

Signals can also fail to propagate if agents direct them based not on their content but on their origin.
Obviously a closed community of agents - one with no arcs to the rest of the graph - will never send any signals to the rest of the graph.
The same thing can happen without an explicitly closed community if agents make some of their output signals insensitive to changes in their input signals.
While this behavior can sometimes be desireable, a ``virtual'' community of agents whose signals to the rest of the network are all insensitive to changes within the community is no different from an explictly closed community.
\label{virtual_community}

%% file: continuity.tex

\section{Separating Propagation from Pattern}
\label{sec:continuity}

I defined separation in Definition \ref{dfn:separation} as the propagation dynamics staying close to an aggregate as that aggregate moves in response to the pattern dynamics.
In this section, I will discuss how stability of the propagation dynamics together with Lipschitz continuity of the pattern dynamics can maintain separation.

The aggregate should not move too fast, compared to the rate at which the propagation dynamics converge.
I will use a differentiable and decreasing function of time $\phi: [0,\infty) \mapsto [0, 1]$ to characterize the rate at which the propagation dynamics converge.
\begin{equation}
  \label{eqn:convergence}
  \phi(t) \rho_E(x_E, p(f_V)) \geq \rho_E(\Gamma^t(f_V, x_E), p(f_V))
  \qquad t\geq0
\end{equation}
Because $\Gamma^0(f_V, x_E) = x_E$, $\phi(0)=1$.
If the propagation dynamics are asymptotically stable, then $\lim_{t \to \infty} \phi(t) = 0$.

Since Definition \ref{dfn:pattern_dynamics} limits the speed at which the agent functions $f_V$ move, the aggregation should also be Lipschitz continuous.
Having supposed a metric $d_v$ on $\F_v$ in Section \ref{agent_metric}, I will use the average of those distances as a metric on $\F_V$.
\begin{equation}
  d_V(f_V, g_V) = \frac{1}{|V|} \sum_{v \in V} d_v(f_v, g_v)
\end{equation}
This particular $d_V$ will make Theorem \ref{thm:Lipschitz_aggregation} slightly easier to prove, but there are other viable choices.
As far as $d_v$ is concerned, a natural choice is:
\begin{equation}
  d_v(f_v, g_v) = \sup_{x_{Tv}}\ \rho_{Fv} \big( f_v(x_{Tv}), g_v(x_{Tv})\big)
\end{equation}

\begin{theorem}
  \label{thm:Lipschitz_aggregation}
  Let the aggregate $p(f_V) = z_E$.
  The aggregation $p$ is Lipschitz continuous if there exists an $M>0$ such that:
  \begin{equation}
    \label{eqn:divergence}
    M (1-\phi(t))\ d_V(f_V, f'_V) \geq
    \rho_E \left(z_E, \Gamma^t \left(f'_V, z_E \right) \right)
    \qquad t\geq0
  \end{equation}
\end{theorem}
\begin{proof}
  Let $z'_E = p(f'_V)$.
  Starting from the triangle inequality:
  \begin{equation*}
    \rho_E \big(z_E, \Gamma^t(f'_V, z_E) \big)
    + \rho_E \big(\Gamma^t(f'_V, z_E), z'_E \big)
    \geq \rho_E \big(z_E, z'_E \big)
  \end{equation*}
  \begin{equation}
    \label{eqn:shifting_aggregate}
    M (1-\phi(t))\ d_V(f_V, f'_V) + \phi(t) \rho_E(z_E, z'_E)
    \geq \rho_E(z_E, z'_E) \qquad t\geq0
  \end{equation}
  This inequality is satisfied trivially at $\phi(0) = 1$.
  At all other points in time,
  \begin{equation*}
    M\ d_V(f_V, f'_V) \geq \rho_E(z_E, z'_E)
  \end{equation*}
  \begin{equation}
    M\ d_V(f_V, f'_V) \geq \rho_E \big(p(f_V), p(f'_V) \big)
  \end{equation}
\end{proof}

Equation \ref{eqn:divergence} limits the rate at which the propagation dynamics, after a change in the agent functions, can diverge from the previous fixed point.
Equation \ref{eqn:shifting_aggregate} pits this rate against the rate of convergence.
It should not actually be necessary for these two rates to cancel out, so long as the rate of convergence exceeds the rate of divergence.

\begin{theorem}
  \label{thm:separation}
  If the aggregate $p$ is Lipschitz continuous and Equation \ref{eqn:convergence} holds, there exists some $t'>0$ such that $\Gamma$ and $\psi_V$ are separate for all $t>t'$.
\end{theorem}
\begin{proof}
  Let $x_E$, $f_V$ be the state of the cognitive network at $t=0$, so $z_E = p(f_V)$ is the initial aggregate.
  Given some small $\delta>0$, let $x'_E = \Gamma^\delta(f_V, x_E)$ and $z'_E = p(\psi_V^\delta(f_V))$.
  Starting again from a triangle inequality:
  \begin{equation*}
    \rho_E(x'_E, z_E) + \rho_E(z_E, z'_E) \geq \rho_E(x'_E, z'_E)
  \end{equation*}
  \begin{equation}
    \label{eqn:induction_step}
    \phi(\delta) \rho_E(x_E, z_E) + M L \delta \geq \rho_E(x'_E, z'_E)
  \end{equation}
  \begin{equation}
    \label{eqn:approach}
    \phi(\delta) + \frac{M L \delta}{\rho_E(x_E, z_E)}
    \geq \frac{\rho_E(x'_E, z'_E)}{\rho_E(x_E, z_E)}
  \end{equation}

  If the right-hand side of Equation \ref{eqn:approach} is greater than one, the signals get further away from the aggregate over the interval $\delta$.
  Less than one, and they get closer.
  If $\rho_E(x_E, z_E) = \epsilon/2$ and the right-hand side is one, then:
  \begin{equation}
    \label{eqn:epsilon}
    \frac{1}{2}\epsilon = \lim_{\delta \to 0} \frac{M L \delta}{1 - \phi(\delta)}
  \end{equation}
  We can ensure this limit exists by requiring that:
  \begin{equation*}
    \frac{d}{dt} M L t \leq \frac{d}{dt} (1 - \phi(t)) \qquad t \in [0, \delta]
  \end{equation*}
  \begin{equation}
    \label{eqn:relative_rates}
    M L \leq -\frac{d\phi}{dt} \qquad t \in [0, \delta]
  \end{equation}
  By decreasing $L$ further, we can make $\epsilon$ arbitrarily small.

  Equation \ref{eqn:induction_step} describes a single step through time.
  If $\rho(x_E, z_E) > \epsilon/2$, then the signals $\Gamma^\delta(f_V, x_E)$ step closer to $p(\psi_V^\delta(f_V))$.
  Adding more steps and taking the limit as $\delta \to 0$, the signals will approach the $(\epsilon/2)$-ball around the aggregate.
  At some finite time $t'$ they will be within $\epsilon$ of the aggregate, and they will never leave that $\epsilon$-ball.
  \begin{equation*}
    \rho_E \Big( \Gamma^t \big(\psi_V^t(f_V), x_E \big), p \big( \psi_V^t(f_V) \big) \Big) \leq \epsilon \qquad t>t'
  \end{equation*}
\end{proof}

As long as the propagation and pattern dynamics are separated, agent $v$ reacts only to $p(\psi_V^t(f_V))_{Tv}$, or at least to arbitrarily similar signals.
If we let $\epsilon$ grow, $\psi_V$ will become stochastic; the propagation dynamics are often stochastic themselves, and there is no way for an agent to know where in the $\epsilon$-ball the aggregate actually is.
It will react to whatever signals it sees.

In fact, the fixed point of propagation dynamics need not be unique.
As long as the propagation dynamics stay near one of the fixed points and the fixed points don't pass within $2\epsilon$ of each other, then the system's choice of a fixed point is unambiguous, and the aggregation is well-defined.
The price we would pay for the extra flexibility is hysteresis in the pattern dynamics.
If multiple fixed points exist, then there is no way to know which one the system will choose without knowing which fixed point it has chosen in the past.

The separation proof can also help us to simulate cognitive networks with a computer.
The constants $M$ and $L$ along with the convergence function $\phi$ imply an upper limit on the time step $\delta$, beyond which Equation \ref{eqn:relative_rates} no longer holds.

%

%% file: signaling_summary.tex

\section{Retrospective on Propagation}

We started with a goal: to design dynamical systems that cognize - that perform the process of knowing.
A cognitive system, simply by obeying its own dynamics, absorbs information and stores that information in its state.
We focused on complex networks because their state space has many dimensions, and high-dimensional spaces can store more information.
But to make full use of that state space, the agents in the network must coordinate with distant counterparts and specialize.

This is quite difficult, because an agent cannot be explicitly aware of any other agents with which it does not share an arc.
Agents may only communicate \emph{through} other agents.
Any change in the agents thus alters their communication channels.

To keep those communication channels stable, we slowed down changes in agent state, giving the propagation dynamics time to converge to a fixed point.
Each component of that fixed point is a signal traveling over an arc in the directed graph.
That single signal can reflect, to varying degrees, the state of every agent.
Because the propagation dynamics are parameterized by the agent states, the fixed point moves as the agents do, and the propagation dynamics never quite reach it.
But the closer they are to their fixed point, the more certain each agent is about the information it receives.
Therefore the slower the agents move relative to the propagation dynamics, the more deterministic the agents' motion.

Although each signal goes from one agent to another, the communication between agents is many-to-many rather than one-to-one.
Every agent sees a few components of the aggregate, as produced by its neighbors.
These were in turn produced from other components of the aggregate, as produced by their neighbors.
The propagation dynamics are the process of aggregation - the process by which the micro level becomes aware of the macro level.

I identify cognition with the emergence of patterns, and the emergence of patterns is widely identified with feedback between the micro and macro levels of a complex network\cite{dewolf_holvoet}.
I have managed to explicitly link the micro level to the macro level - one half of the required feedback process.
To my knowledge, this has never been done before with simple agents and without any omniscient view of the network.

As the agents move in response to that aggregate (at least the components they can see), the aggregate responds to their motion in turn.
The conditions for distributed aggregation are few and broad.
They come in two classes, which I call primary and secondary.
Primary conditions must apply to any network, regardless of the fixed point proof used for the propagation dynamics.
\begin{itemize}[noitemsep]
  \item Agents must be quasi-static on the signaling timescale. ($L$ is small.)
  \item Signals must have a metric ($\rho$).
  \item The propagation dynamics must converge. (At least as fast as $\phi$.)
  \item The propagation dynamics $\Gamma^t(g_V, p(f_V))$ must not diverge too quickly from $p(f_V)$. (Equation \ref{eqn:divergence})
\end{itemize}
The first two conditions apply to the micro-level, which makes them easy to enforce.
While the particular reasons vary from field to field, the heart of the matter is that macro-level behavior is difficult to understand or control.
To try is usually to fail or to diminish the network's cognitive ability.
For instance, what if I were to stipulate that the propagation dynamics contract without requiring that each agent contract?
Some agents would be allowed an $\alpha \geq 1$, but which ones?
It would only take two (mutually adjacent) expansive agents to prevent convergence.
Such a scheme is not impossible - the agents could coordinate their contraction or expansion as part of the pattern dynamics.
But it would be much more difficult to analyze and control.

The second two primary conditions avoid this difficulty through the use of abstraction.
They are metaconditions - conditions about conditions.
Any concrete model of cognition will add its own set of conditions - the secondary conditions.
To give an example, the proofs in Section \ref{sec:stability} both require that each agent contract individually.
That secondary condition need not hold for every cognitive network; the proof of a metacondition can differ for every application.

There is another metacondition hidden in Definition \ref{dfn:propagation}: the requirement that, at a fixed point $z_E$ of $\Gamma_t(f_V, \cdot)$, $z_{Fv} = f_v(z_{Tv})$ for every agent $v$.
It means that the set of potential fixed points is determined on the micro level; the propagation dynamics - a macro phenomenon - only select a fixed point from that set.
The two versions of the propagation dynamics in Section \ref{sec:stability} ($\Gamma^1$ and $\Lambda^t$) actually have the same fixed points.

The three metaconditions I've created so far are relatively easy to tackle because they only restrict the propagation dynamics.
Because the propagation dynamics are often specific to a problem domain anyway, the secondary conditions necessary in the special case sacrifice no descriptive power or ease of application in the general case.
I have striven to keep the my primary conditions as few and as broad as possible.
The agents, for example, are still black boxes.
Their state is not parameterized, which frees us to use nearly any parameterization we like.
They need not even be uniform.
Another example are the signals, which only need a notion of distance.

Most study of complex networks centers on the agents, trying to answer questions like: What do they do? How do they interact? 
My theory shifts focus away from the agents and toward their signals: How do signals propagate? What information do they carry?
Aggregation (as in Definition \ref{dfn:aggregate}) creates a many-to-many communications channel, and Section \ref{sec:continuity} verifies that channel's integrity.
That still leaves us asking what information it conveys.
There are as many answers to this question as there are problem domains, but many domains have commonalities.
For example, an agent that can be connected to an arbitrary number of others must combine an arbitrary number of signals.
A natural way to do this would be to recursively fold pairs of signals together - to give them algebraic structure.
Signal algebras could help determine which information spreads across the network and which does not.

In addition to shifting focus from agents to signals, I have also changed my view of - for lack of a word - \emph{that which cognizes}.
Artificial intelligence researchers usually try to create cognitive algorithms.
Creating an algorithm means breaking down a task or process into discrete steps which mostly occur in linear order.
Aside from the fact that writing a cognitive algorithm has proven to be quite a nasty problem, this approach fosters bad habits of mind.
Because the default point of view is omniscient, we rarely ask key questions about information - how it spreads, how it's stored, where it gets used.
We get used to thinking about information in discrete, hierarchically-organized packets rather than in diffuse gestalts.

Instead of writing cognitive algorithms, I am designing cognitive dynamical systems.
That means accounting for any transmission of information via signaling and limiting each agent's use of information.
It also creates more questions: How do states affect each others' motion?
Can the system's dynamics lead to gestalt configurations of state?
When are those configurations stable?
I will start to answer these questions in Chapter \ref{chp:pattern}.

%% file: pattern_intro.tex

So now I have separated signaling (the propagation dynamics) from the motion of agents (the pattern dynamics), and the agents have time to collectively aggregate macro-level information.
As they move, the aggregation allows them to coordinate their state, together processing information beyond any individual agent's capability.
That information is stored in patterns of agents.
In this chapter I will formally define pattern in a cognitive network, and I will explain how patterns arise from feedback between agent state on the micro level and the aggregate on the macro level.
After that, I will discuss in broad terms how patterns relate to the information they store.

%% file: existing.tex

\section{Existing Approaches to Pattern}
\label{sec:motifs_consensus}

Patterns encode a complex network's knowledge, and their emergence is precisely the process of cognition.
Despite their importance, most fundamental questions about patterns remain unanswered.
When do they arise, and when do they not?
If patterns do arise, which ones?
We may yet answer these questions, but not until we know what a pattern \emph{is}.
I danced around their definition in Chapter \ref{chp:intro}, when I said that patterns are large configurations of agents and their relationships.
These configurations, if the system is to cognize large chunks of information, must be commensurately large, and they must also be unpredictable, stable and coherent.

We have tried to identify the building blocks of patterns.
Network motifs\cite{network_motifs} are small subgraphs with fixed numbers of vertices that repeat across the network.
For a fixed $n \in \N$, there are a limited number of possible $n$-vertex motifs.
The network motif approach starts by identifying, through the use of statistical tests, which three-vertex motifs occur unusually often.
It then moves on to larger motifs, controlling for any smaller motifs already found.
The approach usually stops at five-vertex motifs.
Networks in the same problem domain exhibit similar motifs, which can teach us about the style of patterns in that problem domain.
But if all networks in a domain have the same motifs, then those motifs do not encode an individual network's knowledge.
Rather, networks store information by combining motifs into patterns.

It remains unclear how motifs come to be combined.
It's also not clear that I should be asking how motifs combine instead of asking how agents combine.
To ask how patterns form is to ask how an agent coordinates with distant neighbors of whom it cannot be aware.
I could allow agents to know the motifs of which they are a part - relaxing my simplicity axiom - but I could not explain how motifs coordinate without being aware of each other any more than I could explain how agents do the same.
Furthermore, because I am not working within any particular problem domain, I know less about individual motifs than I know about individual agents.
I conclude that while motifs are a useful tool on the micro-level, shifting focus from agents to motifs would add complication without answering my fundamental questions about cognition and pattern emergence.

Another approach is to form patterns by explicit consensus\footnote{Any kind of pattern will constitute at least an implicit consensus.} of the constituent agents.
There is a large body of work on consensus networks, ably reviewed in \cite{consensus}.
The consensus definition of consensus is ``to reach an agreement regarding a certain quantity of interest that depends on the state of all agents.''
This usually means that ``the state of all agents asymptotically be[comes] the same.''
To the degree that agent states do not become the same, they do not consense.
A pattern does not follow automatically from consensus, but the agents can agree on a pattern to form together.
In this case, the state they agree on must be sufficient to select and describe that pattern.
Because each agent must store this state, we arrive back at a tradeoff similar to the one I described in Section \ref{sec:aggregation}.

That tradeoff pitted the simplicity\footnote{Narrowly: relative to the network as a whole.} of the agents against their need for macro-level information.
I resolved it by using the propagation dynamics to compute an aggregate, thereby distributing the work of aggregation among all the agents.
Consensus networks fail to distribute the storage of information among all the agents, which pits the simplicity of the agents against the network's ability to store information.
In other words, consensus networks must trade off agent simplicity against the network's cognitive ability.
An agent that can - by itself - describe large, coherent patterns is no longer simple relative to the network.
Thus consensus and coherence are at odds, and we find ourselves trading off agent simplicity against the scope and coherence of pattern.

I seek to sidestep this tradeoff as I did the last... or to sidestep the same tradeoff again.
Are these tradeoffs distinct, or are they aspects of the same fundamental paradox?

%% file: emergence.tex

\section{Feedback Loops}
\label{sec:emergence}

It is generally agreed that patterns in a complex network emerge as feedback loops between micro and macro behavior\cite{dewolf_holvoet}.
But what is a ``feedback loop between micro and macro behavior''?
In a cognitive network, the agents (micro level components) collectively compute an aggregate (at the macro level), which in turn determines agent motion (on the micro level).
That interplay creates the possibility for micro-macro feedback, but it doesn't guarantee that feedback occurs.
What would make that guarantee?

I will start with a pair of agents $u,v \in V$.
Each agent has an arc to the other - $(u,v), (v,u) \in E$.
If they coordinate, then $z_{(u,v)}$ should reinforce $z_{(v,u)}$ and vice versa.\footnote{I'm using $z$ rather than $x$ to remind you that agents react the aggregate.}
Reinforcement suggests at least a partial ordering of signals, which I will denote with ``$\prec$''.\footnote{
  If you're still thinking about signals as measureable sets like I suggested in Section \ref{set_signals}, then perhaps for $x,y \in X$, $x \prec y$ if and only if $x \subset y$.
  A stronger ordering might be $x \prec y \iff \mu(x) < \mu(y)$.}
This ordering may vary according to the micro-level context.

Mutual reinforcement also suggests causality.
An increase in $z_{(u,v)}$ should \emph{cause} agent $v$ to change its function $f_v$ in a way that increases $z_{(v,u)}$.
The converse should also be true.
I will treat each agent $v$ as a greedy optimizer of some real function $q_v: X^{|Tv|+|Fv|} \mapsto \R$.
For a positive $\delta \to 0$,
\begin{equation}
  \label{eqn:argmax}
  \psi_v^\delta(f_v) = \argmax_{f'_v \in \F_v}
  \left\{ q_v \big(z_{Tv}, f'_v(z_{Tv}) \big)
  : d_v(f_v, f'_v) \leq L \delta \right\}
\end{equation}
You might protest that making every agent a greedy optimizer is too severe a restriction, that it would at the very least exclude networks made up of people.
But notice that I have not defined $q_v$.
I am not choosing which kinds of agents to analyze; I am abstracting the way \emph{any} agent reacts to its surroundings.
Studying the motion of agents would normally require calculus, which would in turn require that the state spaces $\F_v$ and $X$ be differentiable.
As you will see in Chapter \ref{chp:applications}, differentiability really is too restrictive.
Equation \ref{eqn:argmax} only requires that a utility function $q_v$ exist that agent $v$ greedily optimizes.
Game theory will replace calculus.

Greedy optimization can describe a very wide variety of agents and behaviors.
Agents in a market optimize profit, for instance, and molecules seek their lowest-energy state.
Even in more complicated situations, the formalism can still be useful.
Take the market example.
A person might choose less-profitable work because it lets him or her spend more time with family.
In that case, utility should reflect both profit and family time.
Another might sacrifice short-term earnings in exchange for long-term gains.
We can express such priorities by reformulating the signals as functions of time.\footnote{
  Computer simulation of existing cognitive networks may prove to be impractical.
  There may not exist a good finite approximation of the signals, or the agents' utility functions may be impossible to observe in full.
  The optimization carried out by each agent may defy our power to solve it on a large scale.
  This potential limitation is troubling, but I must ignore it in order to continue.}

For two agents to coordinate, their utility functions must be complementary in the game-theoretic sense.
More simply: each agent must increase the marginal gain its partner reaps by coordinating.
Let $x_E \backslash x'_e = x_{E-\{e\}} \| x'_e$ and more generally $x_A \backslash x'_B = x_{A-B} \| x'_{A \cap B}$.
With that, and given a current aggregate $z_E$, let $x_E, x'_E \in X^{|E|}$ such that $\rho_E(x_E, z_E) < \epsilon$ and $\rho_E(x'_E, z_E) < \epsilon$, where $\epsilon$ is the separation tolerance from Definition \ref{dfn:separation}.
If $x_{(u,v)} \prec x'_{(u,v)}$ and $x_{(v,u)} \prec x'_{(v,u)}$,
\begin{equation}
  \begin{split}
    q_v \big(x_{Tv} \backslash x'_{(u,v)}, x_{Fv} \backslash x'_{(v,u)} \big)
    &- q_v \big(x_{Tv} \backslash x'_{(u,v)}, x_{Fv} \big)
    \\ &\geq q_v \big(x_{Tv}, x_{Fv} \backslash x'_{(v,u)} \big)
    - q_v \big(x_{Tv}, x_{Fv} \big)
  \end{split}
  \label{eqn:neighborv}
\end{equation}
\begin{equation}
  \begin{split}
    q_u \big(x_{Tu} \backslash x'_{(v,u)}, x_{Fu} \backslash x'_{(u,v)} \big)
    &- q_u \big(x_{Tu} \backslash x'_{(v,u)}, x_{Fu} \big)
    \\ &\geq q_u \big(x_{Tu}, x_{Fu} \backslash x'_{(u,v)} \big)
    - q_u(x_{Tu}, x_{Fu})
  \end{split}
  \label{eqn:neighboru}
\end{equation}
This condition - called supermodularity - will not always lead the signals $z_{(u,v)}$ and $z_{(v,u)}$ to both increase or both decrease.
It may be that no $f_v \in \F_v$ exists that increases $z_{(v,u)}$, in which case agent $v$ would like to continue the feedback process but cannot.
The same could also be true for agent $u$.
Nevertheless, supermodularity is necessary for positive feedback between our two agents.

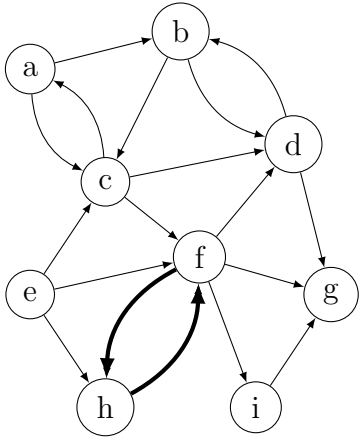
\begin{figure}
  \centering
  \begin{tikzpicture}
    \pic at (0,0) {graph};
    \draw [->,ultra thick] (h) to [bend right=30] (f);
    \draw [->,ultra thick] (f) to [bend right=30] (h);
  \end{tikzpicture}
  \caption[Feedback between adjacent agents]{Feedback between agents `h' and `f'.}  
\end{figure}

Many pairs of agents are not mutually connected.
Even if they are, they may be involved in a larger feedback process that excludes the arcs that connect them directly.
Distant agents communicate through chains of adjacent agents.
A walk should therefore exist from each member of a pair to its counterpart.
Concatenating these walks makes a single closed walk.
Because a closed walk can start from any of its vertices and end up back where it started, it can be expressed as a subgraph - rather than as a sequence of alternating vertices and arcs - without losing any mathematical structure.
Thus we can say that for two agents $u$ and $v$ to coordinate, there must exist some closed walk $C = (V_C, E_C)$ such that $u,v \in V_C$.
In the case of Equations \ref{eqn:neighborv} and \ref{eqn:neighboru}, $V_C = \{u,v\}$ and $E_C = \{(u,v), (v,u)\}$.

\begin{figure}
  \label{fig:feedback}
  \centering
  \begin{tikzpicture}
    \pic at (0,0) {graph};
    \draw [->,ultra thick] (h) to [bend right=30] (f);
    \draw [->,ultra thick] (f) to [bend right=30] (h);
    \draw [->,ultra thick] (f) to (d);
    \draw [->,ultra thick] (d) to [bend right=30] (b);
    \draw [->,ultra thick] (b) to (c);
    \draw [->,ultra thick] (c) to (f);
  \end{tikzpicture}
  \caption[Feedback between distant agents]{Feedback between (say) agents `h' and `d'. Agents `b', `c', `f' and `h' are equal participants in the feedback loop.}
\end{figure}
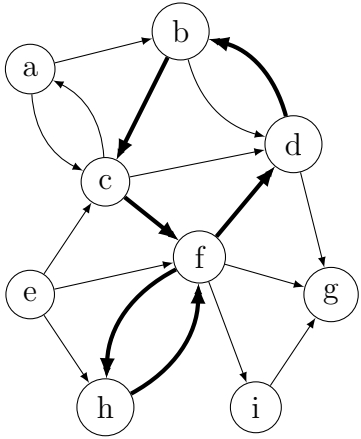

To say that the signals between $u$ and $v$ reinforce each other is to say that the signals along an entire closed walk reinforce each other.
It follows that if this pair of distant agents coordinates, so too must every agent in some closed walk that includes the pair.
That closed walk could include any number of walks between $u$ and $v$, so long as the entire walk keeps reinforcing itself.
In fact, $C$ could be any strongly connected subgraph of $G$. 
Remember that each agent $v$ is only explicitly aware of itself and the signals $z_{Tv}$.
All distant agents are anonymous and also of indeterminate number and connectivity.
Therefore any feedback process between distant agents is a micro-macro feedback loop.
Patterns arise from positive feedback over closed walks in the graph.

\begin{define}
  \label{dfn:feedback}
  Let $C=(V_C,E_C)$ be a closed walk in $G$ and $z_E = p(f_V)$.
  Let the pattern and propagation dynamics be separated by $\epsilon$.
  Take any $x_E, x'_E \in X^{|E|}$ such that $\rho_E(x_E, z_E) < \epsilon$ and $\rho_E(x'_E, z_E) < \epsilon$.
  $x_e \prec x'_e$ for all $e \in E_C$.
  $C$ is a \textbf{feedback loop} if, for all $v \in V_C$,
  \begin{equation}
    \label{eqn:feedback}
    \begin{split}
      q_v \big(x_{Tv} \backslash x'_{E_C}, x_{Fv} \backslash x'_{E_C} \big)
      &- q_v \big(x_{Tv} \backslash x'_{E_C}, x_{Fv} \big) \\
      &\geq q_v \big(x_{Tv}, x_{Fv} \backslash x'_{E_C} \big)
      - q_v(x_{Tv}, x_{Fv})
    \end{split}
  \end{equation}
\end{define}

\begin{define}
  \label{dfn:robust}
  A feedback loop is \textbf{robust} if Equation \ref{eqn:feedback} holds for any $f_{V-V_C} \in \F_{V-V_C}$.
\end{define}

I define robustness here mostly to point out that the agents $V-V_C$ can affect the feedback condition in Equation \ref{eqn:feedback}.
There are likely many degrees and kinds of robustness, but that analysis is out of my scope.

The union of contiguous closed walks is also a closed walk, and the same is true for unions of contiguous feedback loops.
If a feedback loop breaks, parts of it may either remain or not.

\begin{define}
  \label{dfn:coherence}
  A feedback loop is \textbf{coherent} if it has no subgraphs that are feedback loops in their own right.
\end{define}

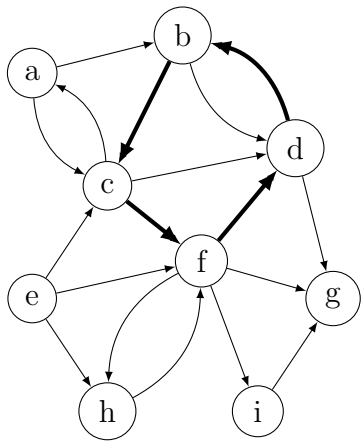
\begin{figure}
  \centering
  \begin{tikzpicture}
    \pic at (0,0) {graph};
    \draw [->,ultra thick] (f) to (d);
    \draw [->,ultra thick] (d) to [bend right=30] (b);
    \draw [->,ultra thick] (b) to (c);
    \draw [->,ultra thick] (c) to (f);
  \end{tikzpicture}
  \caption[A coherent feedback loop]{Depending on the feedback condition for agent `f', agent `h' may not be necessary to maintain feedback in the rest of the loop. The coherent feedback loop (with `h' removed) is pictured.}
\end{figure}

A feedback loop that is not coherent may include virtual communities of the kind I mentioned briefly in Section \ref{virtual_community}.

\begin{define}
  \label{dfn:pattern}
  A \textbf{pattern} is a feedback loop in which the aggregate subvector $z_{E_C}$ is Lyapunov stable.
\end{define}

Definition \ref{dfn:pattern} is similar to a Nash equilibrium\cite{games}, but it does not require the entire system to be stable.
Even the agents $V_C$ in the pattern can continue to move, so long as the aggregate $z_{E_C}$ remains at rest.
But how does the aggregate become stable if it undergoes positive feedback?
A runaway feedback loop will not reach a steady state.
The answer is to bound the image of the aggregation $p(\F_V) \subset X^{|E|}$.
Any positive feedback loop in a compact state space will either reach those bounds or break its own supermodularity condition first.
\begin{figure}
  \centering
  \begin{tikzpicture}
    \draw [fill=gray!60!white] (0,0) ellipse [x radius=1.25cm,y radius=2.5cm]
      node {$\F_V$};
    \draw [fill=gray!30!white] (5,0) ellipse [x radius=1.5cm,y radius=3cm];
    \node (x) at (6.5,0) [right] {$X^{|E|}$};
    \draw [fill=white] (4.6,0) ellipse [x radius=1cm,y radius=2cm]
      node {$p(\F_V)$};
    \draw (0,2.5) -- (4.6,2) node [midway,above] {$p$};
    \draw (0,-2.5) -- (4.6,-2);
  \end{tikzpicture}
  \caption[Bounding the image of the aggregation]{Bounding the image of the aggregation $p$. Return to Section \ref{sec:aggregation} for details about the aggregation.}
\end{figure}
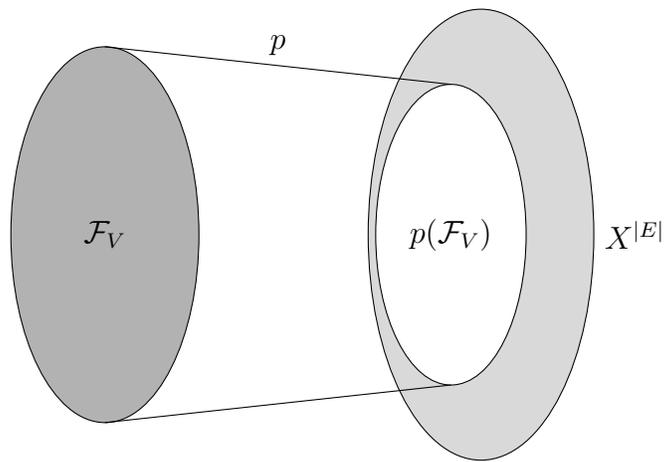
Since a pattern is a stable feedback loop, it must always lie on a bound of the aggregate space.
In fact, the pattern could not be stable without the feedback loop holding it against the boundary.
The image of the aggregation can be manipulated via the agent function spaces $\F_v$.

%% file: specifications.tex

\section{Coherence and Causality}
\label{sec:specifications}

Let us return to the specifications we set for pattern in the introduction: stability, coherence and unpredictability.
Patterns are stable by definition.
Some are coherent.
I will argue that the definition of coherence in Section \ref{sec:emergence} is a good one, and that coherence implies unpredictability.

If cognition is the process by which a system knows or is aware of something, then patterns embody and encode that knowledge.
I said in Chapter \ref{chp:intro} that a pattern is coherent if every part of it depends on the rest.
By Definitions \ref{dfn:feedback}, \ref{dfn:coherence} and \ref{dfn:pattern}, every $z_e, e \in E_C$ can depend in some way on every state $f_v, v \in V_C$.
Each agent's role in the pattern is defined by its neighbors, whose roles are in turn defined by their neighbors.
A pattern, being a \emph{stable} configuration of agent \emph{relationships}, constitutes an implicit consensus of the agents $V_C$.
Rather than rule out explicit consensus, Definition \ref{dfn:pattern} renders it a trivial special case.\footnote{See Section \ref{sec:aco} for an example.}

In general, a pattern encodes knowledge that no individual agent or motif can.
Furthermore, because no agent could produce the stable $z_{Fv \cap E_C}$ without the proper $z_{Tv \cap E_C}$, no one agent or relationship encodes any independently useful knowledge.
Instead, the smallest useful unit of knowledge is the coherent pattern.
Not all patterns are coherent; a pattern can be composed of smaller, contiguous patterns.
But if a pattern is coherent, then any closed loops within it depend on the rest of the pattern in the same way that individual agents do.

Where does the knowledge in a coherent pattern come from?
I will examine the various sources of information which a network may cognize: initial state, the system's own dynamics and the environment.
Each of them will influence most patterns in most networks to some degree.

\label{absorb_environment}
The environment is the most obvious of the three influences.
By the second law of thermodynamics, no living network can subsist without drawing order from its environment.
Despite the close ties between a cognitive network and its environment, I have not yet specified how the two are related.
Let the graph $G$ have at least one special vertex $w$, which represents the environment rather than an agent.
Any signals coming from this vertex are said to enter the network, and any signals sent to it are said to leave the network.
In order for feedback loops including $w$ to stabilize into patterns, the function $f_w$ must be quasi-static on pattern timescales.
Since $w$ has no utility function, the definition of a feedback loop must exclude it.
Making the environment quasi-static is not as restrictive as it sounds, for two reasons.
First: signals can represent probability distributions or cycles of interactions over time.
Second: feedback relations with the environment can be modeled by simply giving $w$ a utility function and allowing it to move, thereby folding it into the model.

Some of a network's knowledge comes from its own dynamics.
It will tend to form certain patterns and not others, based on the objective functions $q_V$ and the search spaces $\F_V$.
Some aspects of this style can be captured by network motifs.
Insofar as motifs are common across many instances of a network model, they isolate style from environmental influence or prior information.
Of course the details of a network need not repeat to be cognized - feedback loops can amplify the information available to even one agent.
These non-repeating details will not affect the network's motifs.

Lastly, patterns can reflect the network's initial state.
They usually will, unless there is only one viable pattern in the whole state space.
Initial states may exist which do not lead to a pattern, even if viable patterns exist somewhere in the state space.
An engineer can prevent this case by ensuring that the pattern dynamics are topologically mixing in the case that no pattern arises.
In the search for a pattern, the system should eventually visit any particular region of its state space.

If a coherent pattern reflects knowledge of all three influences above - initial state, the environment and the system's own dynamics - then these influences are interwoven and cannot be isolated from one another.
Our inability to correspond subsets of knowledge with subgraphs of a coherent pattern prevents us from predicting coherent patterns.
Specifically, it disrupts the causal relationships between patterns (effects) and the conditions from which they arise (causes).
We cannot know which pattern will arise from any given set of conditions, and any given feedback loop could have begun in any number of ways.
A system in which cause and effect cannot be linked is acausal.
This is curious because - assuming the pattern and propagation dynamics are perfectly separated - the pattern dynamics are both acausal and deterministic.

In Chapter \ref{chp:pattern}, I have continued to shift focus away from agents and towards signals.
I defined feedback loops and patterns not in terms of the agent states but in terms of the signals between the agents.
The signals in a pattern must be stable, yet the stability of the agents in the pattern is unimportant in and of itself.
And despite the focus on signals, my theory derives much of its power from keeping those signals unrestricted.
Formally, they must have a metric and a partial ordering - nothing more.
The partial ordering can even be unique to each pair of agents (with only minor changes to Section \ref{sec:emergence}).
Every concrete example in Chapter \ref{chp:applications} will take advantage of the flexibile definition of a signal.

%% file: applications.tex

This chapter will explain how various domains of study fit into our new theory of cognition.
Its purpose is to better explain cognition by providing exemplars, rather than to learn about those exemplars per se.
Most formal proofs are also out of scope - a matter for future work.
That said, I expect that developing these applications further will bear fruit in their respective domains.

Each section will begin with a short overview of the exemplar's model before drawing an analogy between that model and my theory of cognition.
I will start off with Prigogine's dissipative structures.\cite{dissipative_structure}
They will help me introduce a concept common to many applications: the two-way networks.
Unfortunately, a formal theory of two-way networks must wait for another time.

Where dissipative structures are physical systems, cognition can also apply to information processing.
Artificial neural networks will serve as an exemplar.
They will be followed by market economies and then by ant colony optimization.
Cognition should apply naturally to ecology as well, but I know too little of that field to make the attempt.



%% file: autocatalysis.tex

\section{Dissipative Structure}
\label{sec:autocatalysis}


As I said in Section \ref{absorb_environment}, many cognitive networks absorb their knowledge from their environment.
Often, that knowledge reflects two aspects of the environment that are in disequilibrium, by which I mean that the environment has a high degree of order derived from the separation of those two aspects.
Knowing the disequilibrium allows the cognitive network to alleviate it by bridging the two aspects, transporting or transforming \emph{something} from one aspect to the other.
That \emph{something}, which I will call the ``throughput'', may be matter or or information or anything else that can be expressed as a measurable set.
Autocatalytic reaction-diffusion systems are a simple example of a two-way network.

A reaction-diffusion system is a set of reactions occurring in a set of places.
For our purposes, each reaction in a particular place is an agent, and sets of reactants make up the signals.
The products of each reaction either react again in the same place, diffuse to another place, or leave the system.
Alan Turing and Ilya Prigogine analyzed reaction-diffusion systems that produce their own catalysts\cite{morphogenesis}\cite{dissipative_structure}.
They found that when concentrations of certain reactants are kept high and the concentrations of certain products are kept low, the reaction-diffusion system produces its own catalysts so as to increase the rate at which the abundant reactants are transformed into the scarce products.
This is autocatalysis.
The newly-catalyzed reaction produces entropy in the throughput faster than before.
At the same time, the entropy in the system's state decreases.
The second law of thermodynamics remains unbroken because the system produces more entropy in the throughput than it loses by cognizing the disequilibrium.
Eventually the system stablilizes with high concentrations of the catalysts, far from the thermodynamic equilibrium.

An autocatalytic loop is a feedback loop - a closed walk in the graph of reactions over which an increase in each reaction's rate causes the rate of subsequent reactions to increase.
Through these feedback loops, the reaction-diffision system cognizes the chemical potential between certain of its original reactants and ultimate products.
When the feedback stabilizes, the system has stored information about that chemical potential as a pattern of reactions and the reactants they exchange.

I have just described the pattern dynamics of a reaction-diffusion system.
They looked straighforward; what then makes the reaction-diffusion system a two-way network?
The answer has to do with the flow of information.
By the law of mass action, rates of reaction depend on the concentrations of both reactants and products.
As the reactants travel ``forward'', information about the products must travel ``backward''.
On fast timescales - while the various chemical concentrations are quasi-static - the flow of information forward and backward works itself out into deterministic rates of reaction.
The propagation dynamics are the process by which this two-way flow stabilizes.

Let us call the space of forward signals $X$ and the space of backward signals $Y$.
Each type of signal in a two-way network travels over its own graph.
The two graphs are mirror images - wherever one has an arc $(u,v)$, the other has an arc $(v,u)$.
The vertex set is the same for both graphs.
Because each graph implies the other, we can still talk about a single graph $G$ by choosing a convention.
Agent functions map both kinds of inputs to both kinds of outputs.
\begin{equation}
  f_v: X^{|Tv|} \times Y^{|Fv|} \mapsto X^{|Fv|} \times Y^{|Tv|}
\end{equation}
Proving that the propagation dynamics contract becomes more complicated than in Section \ref{sec:stability}, and not just because both kinds of signal must converge.
Specifically, there can be no positive feedback between the two types of signal on those timescales.
Positive feedback would create acausality\footnote{
  I argued in Section \ref{sec:specifications} that pattern emergence is acausal.}
  where the fixed point of the propagation dynamics should be causal.
If the relationship between agent states $f_V$ and the signals $X^{|E|}$ were acausal, the aggregation $p: \F_V \mapsto X^{|E|} \times Y^{|E|}$ could not exist.

As I have labeled the flow of reactants as forward and the flow of products as backward, I will label the source of reactants as the ``entry'' and the sink of products as the ``exit''.
There is nothing special mathematically about forward signals or backward signals; these labels are a matter of convention.
The entry and exit are special vertices in the graph representing the environment.
They may even be the same vertex - $w$ - that I introduced in Section \ref{sec:specifications}.
Like I discussed in that section, $f_w$ should not change over time.

Turing and Prigogine, in their work on what Prigogine termed dissipative structures, identified some physical mechanics of cognition.
But these physical mechanics do not describe all cognitive systems.
Most living things, for instance, are cognitive networks in which each agent is itself a cognitive network.
Obviously these cannot be modeled using reaction-diffusion equations.
Cognition is more an abstract pattern than a concrete phenomenon.

Turing and Prigogine's work was also limited by the need to write down all the equations of motion.
Doing this for large networks or interlocking feedback loops is impractical.
Adrian Bejan sidestepped this limitation in his related constructal law\cite{bejan}, but he substituted another: constructal theory is limited to trees (as opposed to graphs).
Bejan's theory was also limited by its lack of formality.

%% file: neural_nets.tex

\section{Artificial Neural Networks}
\label{sec:neural_nets}

Artificial neural networks are a computational and mathematical abstraction of biological neurons\cite{nnets_history}.
The ``neurons'' are arranged in a directed graph.
At any particular time, every neuron is either firing or inactive.
It activates if the weighted sum of other neurons' activity passes a threshold\cite{mcp}.
Let $x \in X = \{0,1\}^M$ be a neuron's firing activity for $M$ data points.
For an agent $v$ with a bias $b_v$ and a weight $w$ for every input link,
\begin{equation}
  \label{eqn:linear_threshold}
  x_j = H \left(b_v + \sum_{i \in Tv} w_i x_i \right) \qquad j \in Fv
\end{equation}
where $H$ is the Heaviside step function.
To use an artificial neural network, one sets the firing activity of a subset of neurons (called the input neurons), waits for changes in firing activity to propagate through the network and, finally, reads the firing activity of another subset of neurons (the output neurons).
The mapping between input and output for the network is thus determined by:
\begin{itemize}[noitemsep]
  \item the graph
  \item the weights of connections between neurons
  \item the neuron biases
  \item the sequence in which neurons update their firing activity
\end{itemize}
Researchers have studied many, many ways to connect the neurons, sequence their updates and train the weights and biases.
The most popular schemes find ways to serialize the process (making the computation easier) and keep it predictable.

One such scheme is the multilayer perceptron with ``vanilla'' backpropagation.
This scheme is a two-way cognitive network in which the neurons are agents.
In a multilayer perceptron, the neurons are arranged in $K$ layers, where $L_k \subset V$ for each $k \in \{1, \ldots, K \}$.
The input neurons $L_1$ are the entry vertices.
To fit my model, an exit vertex $v'$ must be added to the perceptron; that exit vertex has one arc from each neuron in the output layer $L_K$.
\begin{align}
  &V = \left(\bigcup_{k \leq K} L_k \right) \cup \{v'\} \\
  &E = \left(\bigcup_{k < K} \{(u,v):u \in L_k, v \in L_{k+1}\} \right) \cup \{(u,v'):u \in L_K\}
\end{align}
Each neuron has an arc from every neuron in the previous layer and an arc to every neuron in the next layer.

Since the neurons have no other connections, the firing activity for each data pattern can be computed layer-by-layer with increasing $k$, and each neuron need only update once for the forward aspect of the propagation dynamics to converge.
This layered graph also allows an error gradient to be computed for each weight and bias and data point, using the chain rule to propagate partial derivatives backward from the outputs\cite{backprop}.
Of course, the activation function must be softened to a sigmoid rather than a hard threshold, and each signal $x \in X$ becomes a vector of $M$ reals.
This softening does not affect the network's computational ability\cite{graded_response}.
In the equations for $f_v$ ($v \ne v'$ and $v \notin L_1$) below, $S(x)$ is the logistic function.
Equation \ref{eqn:forward} computes firing activity and Equation \ref{eqn:backward} computes gradients.
\begin{align}
  &x_j = S \left(b_v + \sum_{i \in Tv} w_i x_i \right) \qquad j \in Fv \label{eqn:forward} \\
  &y_i = w_i \cdot \frac{dS}{dx} \left(b_v + \sum_{i \in Tv} w_i x_i \right) \cdot \sum_{j \in Fv} y_j \qquad i \in Tv \label{eqn:backward}
\end{align}
Just like with forward propagation, each neuron need only compute partial derivatives once.
The state space of each hidden neuron $\F_v$ (again, $v \ne v'$ and $v \notin L_1$) corresponds one-to-one with $\R^{|Tv|+1}$ - the space of its input weights and its bias.
The firing activity of each input neuron $v \in L_1$ is set by the environment and does not change.
The gradients from $v'$ are computed by comparing each $x_i, i \in Tv'$ to a desired firing activity $x^d_i, i \in Tv'$ from the training set.
They are then backpropagated using Equation \ref{eqn:backward}.
\begin{equation}
  \label{eqn:error_gradient}
  y_i = x^d_i - x_i \qquad i \in Tv'
\end{equation}
This is the gradient of a squared error function:
\begin{equation}
  \label{eqn:error}
  Err = \frac{1}{2} \sum_{i \in Tv'} y_i^2
\end{equation}

The total propagation dynamics $\Gamma^1: \F_V \times X^{|E|} \mapsto X^{|E|}$ consist of evaluating Equation \ref{eqn:forward} for each neuron in $L_1$ and then $L_2$ and so on until $L_K$, at which point $v'$ computes error gradients using Equation \ref{eqn:error}, each neuron in $L_K$ computes error gradients from Equation \ref{eqn:backward}, $L_{K-1}$ does the same, and so on until $L_2$.
The input layer $L_1$ has no need of error gradients.

The pattern dynamics $\psi_V^1:\F_V \mapsto \F_V$ consist of each neuron greedily minimizing error by descending the error gradient in (ostensibly) small steps.
The gradient for a bias $b_v$ is:
\begin{equation}
  \label{eqn:bias_gradient}
  \frac{\partial Err}{\partial b_v} = \frac{dS}{dx} \left(b_v + \sum_{i \in Tv} w_i x_i \right) \cdot \sum_{j \in Fv} y_j
\end{equation}
The gradient for a weight $w_e, e \in Tv$ is:
\begin{equation}
  \label{eqn:weight_gradient}
  \frac{\partial Err}{\partial w_e} = \frac{\partial Err}{\partial b_v} \cdot y_e
\end{equation}

Here is the objective function for each neuron.
Again, there are $M$ data points, and $m$ is an index that selects a data point.
\begin{equation}
  \label{eqn:learning}
  q_v = -\sum_{m \leq M} \left( \left| \frac{\partial Err}{\partial b_v} \right|_m +
  \sum_{i \in Tv} \left| \frac{\partial Err}{\partial w_e} \right|_m \right) \qquad v \ne v' \text{ and } v \notin L_1
\end{equation}
The change in each weight or bias is the product of some step size $\eta>0$ and the negation of the gradient for that parameter.
Unfortunately the gradients generally become unbounded as they propagate backwards - due to repeated multiplication by numbers potentially greater than one - so there exists no Lipschitz constant $L$ that limits the rate of change of each $f_v$.
Because the propagation dynamics are discrete, this violation of Equation \ref{eqn:argmax} threatens the separation between propagation and pattern dynamics.
This violation is usually refered to as the ``exploding gradients'' problem.

If a weight increases as the pattern dynamics progress, the receiving neuron becomes more sensitive to the firing of the sending neuron.
Since the weight increased in order to decrease the receiver's error, weights of the receiver's outgoing arcs may increase in turn, as the receiver's firing activity has become more valuable.
At the same time, the sender becomes more sensitive to gradients sent back from the receiver.
The sender may henceforth adjust its activity to be of more use the receiver.
These relationships are often crudely summarized as "Neurons that fire together, wire together."
If these relationships hold along a walk from the input layer to the output layer, the corresponding round trip - forward over one graph and backward over the other - constitutes a feedback loop.
A feedback loop becomes a pattern when the gradients along the loop become zero, which happens at a local minimum in $q_v$.

Feedback relations in a neural network are difficult to formulate.
For backward signals, $y \prec y' \iff \sum_{m \leq M} |y| < \sum_{m \leq M} |y'|$.
For the forward signals in a link $e \in E$, $x_e \prec x_e'$ iff $x'_e$ decreases the gradient $y_e$ relative to $x_e$.
Unfortunately, there is no way express the forward ordering using only micro-level information.
The effect of a change in firing activity must be propagated all the way to $v'$ and back, so the forward ordering would be recursive and span entire layers of the network.
In simple terms: no neuron actually knows what it wants; it only sees a gradient that points it in a (hopefully) good direction through $\F_v$.
Thus there is no closed-form way to verify Equation \ref{eqn:feedback} for a multilayer perceptron.
If there was, I conjecture that the proof would start from Equation \ref{eqn:neural_feedback} - a version of Equation \ref{eqn:feedback} modified for two-way cognitive networks like this:
\begin{equation}
  \label{eqn:neural_feedback}
  \begin{multlined}
    q_v \left( x_{Tv} \backslash x'_{E_C}, y_{Fv} \backslash y'_{E_C},
     x_{Fv} \backslash x'_{E_C}, y_{Tv} \backslash y'_{E_C} \right) -
    q_v \left( x_{Tv} \backslash x'_{E_C}, y_{Fv} \backslash y'_{E_C},
     x_{Fv}, y_{Tv} \right) \\
    \geq q_v \left( x_{Tv}, y_{Fv},
     x_{Fv} \backslash x'_{E_C}, y_{Tv} \backslash y'_{E_C} \right) -
    q_v \left( x_{Tv}, y_{Fv}, x_{Fv}, y_{Tv} \right)
  \end{multlined}
\end{equation}
In English, this equation states that the neuron $v$'s utility becomes more sensitive to the output signals $x_{Fv \cap E_C}$ and $y_{Tv \cap E_C}$ as the input signals $x_{Tv \cap E_C}$ and $y_{Fv \cap E_C}$ strengthen.
I am not yet sure that Equation \ref{eqn:neural_feedback} is properly formulated.
Such a formulation would be core to a  formal theory of two-way cognitive networks.

Not all walks from input to output are feedback loops.
A marginal improvement may not be enough to make the receiver (the one described above) more valuable than the other neurons in its layer.
And the sender's increased sensitivity to the receiver's gradients may not surpass the sender's sensitivity to the gradients of other neurons in the receiver's layer.
To identify the selection of closed walks that undergo positive feedback, Equations \ref{eqn:forward}, \ref{eqn:backward} and \ref{eqn:error_gradient} should be plugged into a feedback condition like Equation \ref{eqn:neural_feedback}.
Of course, it will be necessary to modify Equations \ref{eqn:backward} and \ref{eqn:error_gradient} so that sensitivity of the backward signals to the forward signals in a link (on pattern timescales) could be expressed in closed form.
This method of analyzing neural learning rules could prove quite useful in future work.

Neurons train without any knowledge of the error itself.
They only see the error gradient.
By descending that gradient only in small steps, the weights and biases try to be quasi-static on signaling timescales.
Unfortunately, the network runs into trouble here.
The gradients in a given layer are roughly proportional to the weights in layers closer to the exit.
Positive feedback loops can cause the gradients to increase exponentially - to ``explode'' - as the backpropagate.
Because the steps taken by the weights and biases are proportional to the gradient, the pattern dynamics can violate their own Definition \ref{dfn:pattern_dynamics} and thus the separation between propagation and pattern.

Just like positive feedback leads to exploding gradients, negative feedback can lead to ``fading'' gradients.
As I warned at the end of Section \ref{sec:stability}, overly-aggressive contraction can prevent signals from spreading properly across the network.
Also, small gradients lead to only small changes in weights and biases, eventually halting the pattern dynamics.

Multilayer perceptrons largely try to avoid problems with exploding and fading gradients by keeping the graph feedforward with few layers.
If each component of the gradient only makes two or three jumps backward, it can't grow or shrink much.
Gradient descent causes more problems in networks with many layers, like deep belief networks.
The problem is worst in recurrent networks, which have directed cycles.
A signal in a recurrent network may make an infinite number of jumps.
If the gradients explode, then the propagation dynamics of such a network will never converge, and without convergence, the aggregation $p$ does not exist.
A gentle fading will make the propagation dynamics contract, but excessive fading still prevents signal propagation.

In contrast, the forward signals do contract.
Although I will not prove this statement, Hopfield provided good inductive evidence\cite{hopfield} that the propagation of firing activity mostly converges.
Intuitively, a neuron rejects any changes in its input that do not push it over its threshold - from inactive to firing or vice versa.


%% file: markets.tex

\section{Market Economies}

Most naturally-occuring cognitive networks are composed of, intertwined with and assemble into other cognitive networks.
Modeling such systems is tricky and intellectually risky.
The modeler must pick and choose what to represent, and any omissions are bound to be criticized by those with different priorities.
In this section, I run a greater risk than most: I am not an economist.
Remember that my purpose is to explain cognition, not the finer points of economics.
The broad pattern of an economy will suffice.
Among the many contexts in which it can be understood, I will model the specialization of market niches and the relationships between them.
I will try to describe how niches develop and coordinate to best acquire, process and allocate resources.

Every model defines a context within the heterarchy of cognitive networks - the heterarchy that extends from individual cells to the biosphere.
A theory of cognition can guide the creation of models and render them in a common language, enabling us to draw clearer lines between contexts and to avoid using a model outside its context.
The content of signals in the model circumscribes what it can cognize, and the choice of agents determines the model's granularity.
A consequence of acausality is that even a good model may not predict which patterns emerge in its referent.
That does not, however, mean the model is useless.
The model might still predict what kinds of pattern can emerge, as well as the stability or coherence of patterns which have already emerged.
In other words, the model might still predict what can be known or is likely to be known.

Each niche is an agent characterized by:
\begin{itemize}
  \item a product
  \item the producers of that product
  \item the method of production
\end{itemize}
The market economy is a two-way cognitive network.
Goods and services are the forward signals while money and preferences are the backward signals.
By transforming goods and services into other goods and services, niches (as agents) collectively bridge the available resources to consumer demand.
Those available resources are the entry vertices.
Consumers are the exit vertices.

The propagation dynamics are the process by which the market works out the price of each good and the volume bought by each receiving niche.
The niches are quasi-static, which is to say the products, sets of producers and the methods of production are all fixed on short timescales.
This limits each niche's flexibility.
Say, for instance, that the demand for a good increases.
At some point the increase will test the capacity of the producing niche.
The price will rise, decreasing demand.
Conversely, a drop in demand will cause a drop in price, which raises demand.
This negative feedback - driven by profit and limits on flexibility - ensures that the propagation dynamics converge.

The pattern dynamics are the adaptation of goods to fit buyer preferences and the adaptation of the means of production to fit the available resources.
Producers can leave or join niches, or they may invest in or liquidate means of production.
There are two kinds of positive feedback.
The first - analagous to autocatalytic loops - occurs when a supply chain produces some of its own means of production.
The second comes from economies of scale and is analagous to the feedback loops in a multi-layer perceptron.
For many goods, the more that gets produced, the easier it is to produce.
The easier a good is to produce, the more profit that can be made, which draws more people and companies to produce that good.
This kind of feedback loop is a supply chain from raw material to consumer.
Like the feedback loops in an ANN, the backward walk mirrors the forward walk.

In general, goods in a supply chain have uses both inside and outside the supply chain.
As positive feedback progresses, those goods will become more suited to their use inside the supply chain as opposed to outside it.
If the graph can change over time, niches will split and differentiate to take advantage of their own economies of scale.\footnote{I have not yet addressed changes to the graph, but such a study would make for fascinating future work.}
In this way, the supply chain may become an industry.

Like any model, this one leaves a lot of interesting behavior out of context.
Here is a short, non-exhaustive list.
\begin{itemize}[noitemsep]
  \item Modeling niches as agents ignores the differences between large companies, small businesses and individuals.
  \item Modeling the physical environment as the entry ignores environmental damage.
  \item Treating the consumer as the exit node omits anything that alters consumer demand, like culture or advertising.
\end{itemize}
I have also left out the precise definition of a good.
The simplest definition is the set of uses for the good.
To model uncertainty - e.g. risk or asymmetries in information - I could make every signal a probability distribution over the power set of possible uses for a good.\footnote{
  Similar to Eigen and Schuster's treatment of genetic risk\cite{gene_risk}.}
That still leaves out temporal behavior, like borrowing money or training the workforce.
To model those, each good would become a function of its uses over a cycle in time.

That should suffice for an example of cognition.
The rest, I will leave to economists.

%% file: ant_colonies.tex

\section{Ant Colony Optimization}
\label{sec:aco}

Ant colony optimization (ACO) is a class of algorithms for finding a ``short'' path from one vertex in a graph to another\cite{aco_review}.
It may seem very specialized, but many combinatoric optimization problems can be expressed as such a path search.
ACO works by imitating the method by which ants settle on a path between their colony and a food source.
Some number of ants start at the colony and wander through a graph of waypoints, guided toward the food by smell (a search heuristic).
After they reach the food, they retrace their steps back to the colony, laying pheremone uniformly along their path.
Each ant has a fixed amount of pheremone to lay, so it will lay pheremone more densely on shorter paths.
Ants that walk the graph thereafter will favor arcs\footnote{
  or edges, in the undirected case
} with more pheremone.
When these ants lay their pheremones in turn, they will tend to lay it on these arcs, which will draw more ants.
As time goes on, the ants settle on a single short path through the graph, and this path maps to a solution of the combinatoric optimization problem.

An ACO is a two-way cognitive network in which the waypoints are the agents.
An agent also includes the pheremone concentrations of its outgoing arcs.
The location of the colony is the entry vertex, and the location of the food source is the exit vertex.
The forward signal in an arc is a probability distribution over all the forward walks an ant could take from the entry vertex through that arc (before reaching the exit).
The backward signal is the same but from the exit vertex and over the backward graph.
These walks are not uniformly random; ants are guided first by the search heuristic and then by pheremone concentrations.\footnote{
  Some instances of ACO don't use a heuristic. Before pheremones are laid, the ants in those colonies do make uniformly random pathing choices.}
The heuristic - the smell of food - has a fixed value for each arc.
When a single ant walks randomly across the graph, it samples a probability distribution for each arc in its path.

The propagation dynamics consist of ants walking across the network while the pheremone concentrations are quasi-static.
That condition can be achieved by keeping each ant's pheremone payload small compared to the ants' ability to smell.
Most implementations pass a population of ants through the network before laying all their pheremones at once, but this is more a matter of convenience than functionality.
More important is the impracticality of covering the entire support of every path distribution.
We cannot say the propagation dynamics have converged until individual ants cease to shift the colony's estimate of these distributions.
But every path would need to be traversed by at least one ant, and the point of ACO is to be faster than simply trying all the paths.
That contradiction sounds catastrophic, yet ACO works.
Why?

It works because ACO need not maintain perfect separation between propagation and pattern dynamics.
Relaxing separation makes the pattern dynamics stochastic with an increasing variance.
That stochasticity doesn't cripple the algorithm because there are usually many good paths.
ACO does try to improve separation by keeping the number of likely paths small.
First, it prohibits ants from revisiting a vertex before reaching the exit.
Second, the search heuristic attracts the ants to a smaller set of paths, concentrating their probability mass on a subset of its support.
Finally, positive feedback helps minimize the number of likely paths.
As time proceeds, a subset of paths comes to dominate the rest, and then a subset of that subset, and so on.
This serves to shrink the effective support of the path distributions.
The result of these steps is an algorithm that reaches a good solution quickly.
For very hard problems, this is often better than taking a long time to find the best solution.

The pattern dynamics are the process by which pheremone is laid.
Having already described the process from the ants' perspective, here is the waypoints' perspective as agents.
An agent maximizes the number of ants on the outgoing arc with the most ants.\footnote{
  This does not mean sending all ants over the same arc as soon as that arc becomes popular; such a shift would be a discontinuous jump through the agent's state space.}
If an agent increases the number of ants it sends along the (expected) shortest path, the expected path length of its ants will decrease.
That gives the waypoints from which it receives ants reason to favor it, because they have the same incentive.
By this reasoning, the shortest walk through the graph should coincide with the sole feedback loop.
The catch is that pheremone concentration is similar but not identical to the inverse of expected path length.
At a certain point in the feedback, when pheremones have become concentrated enough, the ants will choose well-worn paths simply because they are well-worn.
Like with many stochastic optimization algorithms, one must balance speed of convergence with the quality of the solution.

Pheremones evaporate at a rate proportional to their concentration.
Even if all the ants choose the same path, eventually the rate at which those ants lay pheremone equals the evaporation rate.
The evaporation rate is not necessary for a feedback loop to become a pattern - that loop stabilized as soon as its pheremone levels got high enough that nearly all the ants chose it.
Evaporation is necessary if one wishes ants to explore after a pattern emerges.
Such exploration is only useful if the algorithm is elitist\footnote{
  An elitist algorithm always selects and remembers the best solution found so far.};
 otherwise the other ants will ignore any new paths an explorer finds.
Elitism involves computation external to the network, so it has no analog in a cognitive network.

Unlike the other exemplars I've introduced, ant colonies primarily cognize themselves.
Ants enter the network without external information.
Instead, the information encoded in a pattern reflects only the graph, the lengths of the arcs, the search heuristic and the random choices of the earliest ants.

Waypoints are not usually seen as agents; the ants are.
These ant-agents come to an explicit consensus on the best path.
ACO occupies a curious middle ground between cognitive networks and consensus algorithms.
This dual interpretation possible only because feedback in ACO is winner-take-all.
Seeing it as a consensus algorithm, each agent must know the whole pattern.
Seeing it as a cognitive network, the set of possible patterns is very restricted.
These two apparent problems seem to offset each other.

%% file: meta.tex

I have now outlined a new formal theory of cognition.
It is does not focus on humans, on artificial intelligence or on any other specific domain.
Instead, it lays out the abstract process by which a complex network acquires knowledge.
My theory is still limited to fixed graphs\footnote{
  where the signals or agents states move but not the vertex or arc sets
} with two levels - micro and macro.
A complete theory should describe the dynamics of the graph and explain how an arbitrary number of levels interact when arranged in a heterarchy.


%% file: review.tex

\section{Context in Cognition}

A cognitive system is a network of agents.
These agents signal each other on relatively short timescales and change their state over longer timescales.
For every configuration of agent states, the signal propagation dynamics find a stable fixed point, called the aggregate.
Since signals propagate faster than the agents move, the propagation dynamics closely track this fixed point, which means the agents react only to the aggregate.
In the pattern dynamics, each agent maximizes a utility function of its input and output signals.
This behavior may lead subsets of agents coordinate across closed walks through the graph by reinforcing their signals to each other.
A stronger signal over any arc in the walk causes the receiving agent to strengthen the signal it sends over the next arc.
Eventually this wave of influence returns to strengthen the first signal further, and the closed walk becomes a feedback loop.
If the signals in a feedback loop stablize, the feedback loop becomes a pattern.
The selection of closed walks that undergo feedback and then stabilize into patterns, together with the nature of the signals in the pattern, encode the network's knowledge.
Despite being ad hoc and of arbitrary scale, these encodings are coherent.

So far, I have mostly avoided the use of ``complex'' and ``simple'' as adjectives, because they are only useful in a relative sense.
Recall the (biological) ant colony I sketched on page one.
If I modeled the colony, I would probably say that each ant is ``simple''.
Yet an ant is incredibly complex in its own right.
Its brain is a cognitive network of cells, and those cells are cognitive networks of molecules.
Retreating to the perspective of an ecosystem, the colony itself looks like a simple agent.
Every cognitive system operates in some context or set of contexts.

\begin{quote}
  \label{context}
  A \textbf{context} is defined by a $\sigma$-algebra on a set of things that can be known in that context.
\end{quote}

A context can be very broad but never universal.
Billy Vaughn Koen would say that all knowledge is heuristic, in that it ``provides a plausible aid or direction in the solution of a problem but is in the final analysis unjustified, incapable of justification, and potentially fallible.''\cite{koen}
Since any unit of knowledge is part of some contexts but not others, it follows that objective knowledge is a contradiction.
It also follows that knowledge can only be measured relative to a given context, with that context being the domain of the measure.
Since no context exists which includes all possible knowledge, no objective measure of knowledge exists.

Similar reasoning applies to measures of ``complexity''.
As a field, we have been asking how complex behavior arises from simple parts.
But without an objective notion of complexity, this question loses its meaning.
I will reframe it.
By saying that an agent is simple, I am saying that its context includes the signals it receives (and sends) but not the whole network.
Likewise, the network sees an agent's output signals but not the potentially-complex $\argmax$ problem an agent solves internally.
Thus an agent is both simple and complex relative to the whole network, and the whole network is both simple and complex with respect to each agent.
Every cognitive process implies one or more contexts.
Each context is a $\sigma$-algebra of things the cognitive system can know.

\begin{figure}
  \centering
  \begin{tikzpicture}[scale=1.3]
    \draw[very thick] (0,0) circle [radius=1cm];
    \def\n{5}
    \def\angle{360/\n}
    \foreach \i [evaluate=\i as \slice using \i*\angle] in {1,...,\n}{
      \draw (0,0) -- (\slice:1cm);
    }
  \end{tikzpicture}
  \hspace{2cm}
  \begin{tikzpicture}[scale=0.85]
    \draw[very thick] (0,0) circle [radius=1cm];
    \def\n{5}
    \def\angle{360/\n}
    \foreach \i [evaluate=\i as \context using (\i+.5)*\angle] in {1,...,\n}{
      \draw (\context:1.2cm) circle [radius=.8cm];
    }
  \end{tikzpicture}
  \hspace{2cm}
  \begin{tikzpicture}[scale=0.5]
    \draw[very thick] (0,0) circle [radius=1cm];
    \def\n{5}
    \def\angle{360/\n}
    \foreach \i [evaluate=\i as \context using (\i+.5)*\angle] in {1,...,\n}{
      \draw (\context:1.2cm) circle [radius=2.2cm];
    }
  \end{tikzpicture}
  \caption[Overlapping contexts]{
    \emph{On the left:} A single monolithic context divided into micro contexts.
    \emph{In the center:} Overlapping micro (thin) and macro (bold) contexts in a cognitive network.
    Parts of each context are shared, and others are hidden.
    \emph{On the right:} The micro contexts have subsumed the macro context.
    This could involve a centralized aggregation procedure (as opposed to a propagation dynamics) or an algorithm for the explicit consensus of agents.
  }
\end{figure}
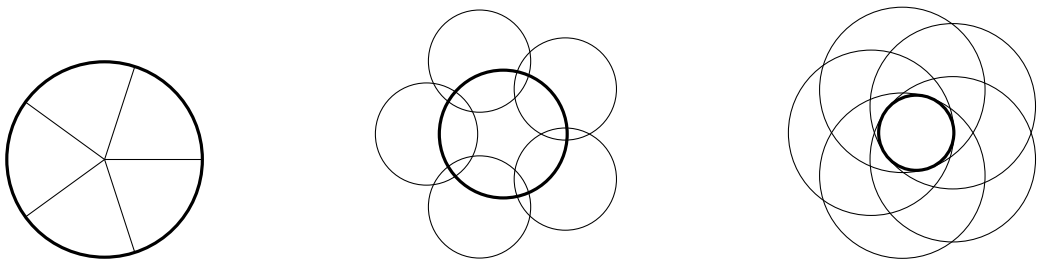

The paradox of ``complex behavior from simple parts'' comes from thinking about a cognitive network in only one context at a time. 
By considering agents in their own micro-level contexts, I have reframed our question: How do the micro and macro contexts of a cognitive network relate to each other?
The macro context consists of feedback loops and patterns, and each micro context consists of an individual agent's utility maximization problem.
My theory links these contexts by means of the aggregation (recall Definition \ref{dfn:aggregate}) and of complementarity between agent utility functions.
This framework begins to explain - formally - how a consciousness relates to the substrate from which it emerges.

The field of cybernetics tried to explain cognition qualitatively.
Maturana and Varela, for example, introduced the concept of autopoiesis, or self-creation, in living systems\cite{autopoiesis}.
Douglas Hofstadter introduced strange loops, which are configurations of symbols that take on circular meaning, thereby becoming self-aware\cite{godel_escher_bach}.
Unfortunately, cybernetics researchers were only able to express their ideas with words rather than with mathematics.
While their work fascinated many, none could agree on what it meant or how it should be applied.
We need formality if we are to reliably convey our understanding of cognition from mind to mind.

Most other work on cognition falls into one of two categories.
The first is a top-down approach that focuses more on \emph{what} a cognitive system does than \emph{how} it cognizes.
Psychology falls into this category, and so does the study of symbolic artificial intelligence.
This approach struggles with context.
Psychological studies are difficult to control because human thought takes place in many contexts, and those contexts are very hard to ascertain or control.
A well-controlled study may yield clear quantitative results, but those results only apply in the narrow context created for the study.
In artificial intelligence, Alan Newell\cite{newell} framed cognition as a rational, goal-oriented search through a problem space.
But a well-posed search calls for a well-defined goal, which cannot be defined a priori for a whole cognitive network.
And rationality - being an application of knowledge - is just as context-dependent as the knowledge itself.
Unfortunately, the context for all known artificial intelligences must either be specified by a human designer or generated by an algorithm written by a human designer.
The context(s) of such a system will always be circumscribed, and the system will always face an intractable and ill-defined meta-problem - figuring out when its knowledge is or is not useful.
A top-down approach requires a clear context that doesn't often exist.

The second category is characterized by a bottom-up approach which seeks the micro-level mechanics of cognition.
This kind of cognitive architecture adapts to the context it finds itself in.
Artifical neural networks are a good example, as are Pentti Kanerva's sparse distributed memories\cite{kanerva}.
Neuroscience searches for the mechanics of human (or animal) cognition.
Like cybernetics, work in this category asks \emph{how} something learns.
But where cybernetics tends to be informal and abstract, bottom-up approaches are formal and concrete.
Every cognitive architecture modeled with a bottom-up approach either arose by itself or explicitly mimicks a system that did.
Until now, no one has been able extract or abstract the essence of cognition for novel use in other systems.

My goal has been a formal, abstract theory of cognition that deals properly with context.
Like the bottom-up approach, I have sought formal mechanics of cognition.
Like cybernetics, I have tried to keep those mechanics abstract.
At the end of Section \ref{sec:motifs_consensus}, I pointed out that the usual tradeoffs between agents and the macro-level -
\begin{enumerate*}[label=\alph*)]
  \item agent simplicity versus the processing of macro-level information and
  \item agent simplicity versus the storage of macro-level information
\end{enumerate*}
- could be two aspects of the same fundamental paradox.
That paradox is how complex behavior arises from simple parts.
I resolved it by finding a better question: Where do the micro and macro contexts overlap, and where do they not?
By treating signal propagation as a dynamical system unto itself, the micro-level contexts aggregate the macro context without subsuming it.
By defining pattern in terms of closed walks of complementarity between agents, the micro-level contexts coordinate to store information in the macro context while remaining distinct from the macro context.

%% file: future.tex

\section{Future Work}

Almost every aspect of my theory could be developed further.

Regarding aggregation, I have not yet explained what information propagates and what doesn't.
Abstract algebra may provide the proper tools to model and control the spread of signals, given how they join and split at each agent.
If signals can be measured, ergodic theory may also be useful.

Two-way networks deserve a formal definition and analysis.
How do forwards and backward signals interact?
What, precisely, constitutes a feedback loop through a pair of mirrored graphs?
Such work will allow us to apply the mathematics of cognition to a much wider range of systems.

The topology of the pattern dynamics is still fairly mysterious.
How robust are patterns, and what kinds of robustness are there?
Can patterns merge or split, and how do these events affect their coherence?
Can parts of a coherent pattern survive if the pattern loses stability?
These questions are fundamental to the design of cognitive networks.

Even though most natural cognitive networks are related heterarchically, my theory analyzes only two levels of behavior at a time - a micro-level and a macro-level.
Can a formal definition of context allows us to model the relationships between many contexts?
An answer to this question is vital for modeling natural systems.

How does the graph of a cognitive network change over time without interfering with separation or breaking feedback loops?
In particular, how can an agent create new arcs when its context is limited to the arcs it already has?
This question is very pertinent to artificial neural networks as well as to design automation.

Finally, there's a lot of application work to do.
Cognition can help design new metaheuristics for optimization, and it might provide a better understanding of the ones we already have - genetic algorithms, ACO and particle swarms in particular.
New ways to train and structure artificial neural networks should present themselves.
And finally, it should be possible to improve the distributed control of large industrial systems like smart grids and reconfigurable manufacturing systems.

Many of these applications will inevitably inform the more abstract inquiries, just like the systems in Chapter \ref{chp:applications} guided me to the theory in Chapters \ref{chp:propagation} and \ref{chp:pattern}.
The social and biological sciences could be a fertile ground for new quantitative models, although we will probably need a formal theory of context in order to comprehensively model market economies or ecologies.
A better understanding of context could likewise provide a formal language for comparing and contrasting cognitive architectures.

Curiosity aside, we should be cautious about modeling social systems.
As we study them, we must remember that social theories can affect their referents if or when they become public knowledge.
These observer effects can be unpredictable or even dangerous.
Given the training necessary to fully understand theories of cognition, these theories are unlikely to find their way (intact) into the public consciousness.
Rogue, poorly applied social theories have in the past created much upheaval.